\documentstyle[preprint,tighten,eqsecnum,floats,epsfig,aps]{revtex}
\font\myit=cmti10

\newcommand{\beq}{\begin{equation}}
\newcommand{\eeq}{\end{equation}}
\newcommand{\mc}{\multicolumn}
\newcommand{\lsim}{\mathrel{\mathop{\kern 0pt \rlap
  {\raise.2ex\hbox{$<$}}}
  \lower.9ex\hbox{\kern-.190em $\sim$}}}
\newcommand{\gsim}{\mathrel{\mathop{\kern 0pt \rlap
  {\raise.2ex\hbox{$>$}}}
  \lower.9ex\hbox{\kern-.190em $\sim$}}}

\begin{document}
\draft

\title{Local Realistic Theories and Quantum Mechanics for the
two--neutral--kaon system}
\author{R. H. Dalitz$^{\mathrm{a}}$ and G. Garbarino$^{\mathrm{a,b}}$}
\address{{\myit $^{\mathrm{a}}$Department of Theoretical Physics, University of Oxford \\
1 Keble Rd, Oxford OX1 3NP, UK} \\
{\myit $^{\mathrm{b}}$Grup de F\'{\i}sica Te\`{o}rica, 
Universitat Aut\`{o}noma de Barcelona \\
08193 Bellaterra, Barcelona, Spain}}
\date{\today}
\maketitle

\begin{abstract}
The predictions of local realistic theories for the 
observables concerning the evolution of a 
$K^0\bar{K}^0$ quantum entangled pair 
(created in the decay of the $\phi$--meson) are discussed. 
It is shown, in agreement with Bell's theorem, 
that the most general local hidden--variable model fails in
reproducing the whole set of quantum--mechanical joint probabilities.
We achieve these conclusion by employing two different approaches.
In a first one the local realistic observables are deduced
from the most general premises concerning locality and realism,
and Bell--like inequalities are not employed.
The other approach makes use of Bell's inequalities.
Within the former scheme, under particular conditions for
the detection times, the discrepancy between
quantum mechanics and local realism for the time--dependent asymmetry
turns out to be not less than 20\%.
The same incompatibility can be made evident by means of a Bell--type test
by employing both Wigner's and (once properly normalized probabilities
are used) Clauser--Holt--Shimony--Holt's inequalities.
Because of the relatively low experimental accuracy,
the data obtained by the CPLEAR collaboration for the asymmetry
parameter do not allow for a decisive test of local realism. Such a test,
both with and without the use of Bell's inequalities, should be feasible in the
future at the Frascati $\Phi$--factory.
\end{abstract}
\pacs{3.65.Bz}

\newpage
\pagestyle{plain}
\baselineskip 16pt
\vskip 48pt

\newpage
\section{Introduction}
\label{intro}
In 1935 Einstein Podolsky and Rosen (EPR in the following) \cite{Ei35} advanced a 
strong criticism concerning the interpretation of quantum theory. 
They arrived at the conclusion that the description of
physical reality given by the quantum wave function is not complete.
EPR's argumentation was based on a condition for a {\it complete theory}
({\it every element of physical reality must have a counterpart in the physical theory})
and on a criterion which defines {\it physical reality}
({\it if, without in any way disturbing a system, we can predict with certainty
(i.e., with probability equal to unity) the value of a physical quantity, then there exists
an element of physical reality corresponding to this quantity}). They
also assumed the quantum world to be {\it local}: this requirement was 
introduced in order to express 
relativistic causality, which prevents any action--at--a--distance.
Starting from these premises,
by considering the behaviour of a correlated and non--interacting
system composed by two separated entities, EPR arrived at the following conclusion:
contrary to what the indetermination principle states,
two non--commuting observables can have simultaneous physical reality,
then the description of physical reality given
by Copenhagen's interpretation, which does not permit such a simultaneous 
reality, is incomplete. 
At the very heart of their logical conclusion is the following
fact: their assumption, according with a quantum system has real and well defined
properties also when does not interact with other systems 
(including a measuring apparatus), is contradicted by quantum mechanics. 

This was the point attached by Bohr in his famous replay \cite{Bo35} to EPR's 
paper. Here he noticed that EPR's criterion of
reality contained an ambiguity if applied to quantum phenomena.
Starting from the complementarity point of view, Bohr stated that {\it quantum
mechanics within its scope {\rm [namely, in its form restricted to human knowledge]}
would appear as a completely rational description of the physical phenomena}.
In the opinion of Bohr the conclusion of EPR was not
justified since they contradicted quantum theory at the beginning, through
their criterion of physical reality: following Copenhagen's interpretation,
quantum reality has to be defined by the experimental observation of phenomena. 

The probabilistic meaning of the quantum wave function is the main
assumption that originated criticisms and debate for
a broader interpretation of quantum theory. In fact, the wave function
provides a description of the microscopic world in accordance with the laws of
chance, namely it is non--deterministic: 
the actual result of a measurement is selected from the set of
possible outcomes at random. It is this interpretation of the quantum state 
that led Einstein to pronounce the historical sentence: {\it God does not play dice}. 

Within Copenhagen's interpretation, the measurement process changes
the state of the measured system through the reduction of the wave packet.
The description of this (non--deterministic and non--local) process given
by the {\it hermitian} operator associated to the observable
one measures is mathematically different from the
(deterministic) evolution of the statistical predictions of
the wave function, which is accounted for by the
Schr\"{o}dinger equation and its {\it unitary} time evolution operator. 
This matter of fact is also the origin of different paradoxical conclusions of quantum
mechanics. It is important to stress that the collapse of the wave function 
is a non--local aspect of quantum mechanics. It
arises from the fact that the theory does not provide
a causal explanation of the anti--correlations which exist between
the probabilities of finding a system (say a particle) in two separated regions of space.
The EPR--type correlations of two--particle entangled states clearly 
exhibits a non--locality. To avoid this feature, interpretations
of quantum mechanics which do not incorporate
the reduction of the wave packet have been introduced
(see for instance Bohmian mechanics \cite{Bo52} and Everett's
many--world interpretation \cite{Ev57}). However, we have to stress that the
non--local features exhibited by EPR's states 
do not contradict the theory of relativity, since they do not allow
for faster--than--light communications \cite{Sh84}.


Another puzzling question concerns the subdivision of
the physical world into {\it quantum system} and
{\it classical apparatus}, the latter being directly controllable and needed to define
(through the measurement process) the properties of quantum phenomena.
Actually, strictly speaking, real physical properties are possessed only by the
combined system of quantum object plus measuring device.
This dualistic approach, which leaves the measuring devices out of the world treated by
the mathematical formalism of the theory, leads to a description of the
physical universe which is not unified, namely to a theoretical framework
which is not fully coherent. 

The first hypothesis for the solution of the paradoxical conclusion of EPR
concerning quantum correlations was proposed by Furry \cite{Fu36} in 1935. He assumed that
the quantum--mechanical description of many--body systems could break--down
when the particles are sufficiently distant one from another
(practically when their wave functions do not overlap any more). This means
that in presence of EPR correlations between two quantum subsystems
which are very far away one from each other, the state
of the global system is no longer given by a superposition
of tensorial products of states but it is simply 
represented by a statistical mixture of products of states (namely it is factorizable). 
However, Furry's hypothesis revealed to be incorrect: an old experiment concerning
polarization properties of correlated photons \cite{Wu50,Bm57}, as well as
more recent tests \cite{Ap98,Ze99,Ti98}, excluded a possible separability of the
many--body wave function even in the case of space--like separated particles. 

In 1952 Bohm \cite{Bo52} suggested an interpretation of quantum theory
in terms of {\it hidden--variables}, in which the general mathematical
formulation and the empirical results of the theory remained unchanged. 
In Bohm's interpretation 
the paradoxical behaviour of correlated
and non--interacting systems revealed by EPR find an explanation. 
However, for such systems Bohm's theory exhibits a
non--local character, which cannot be reconciled with relativity theory.

This result is consistent with what Bell obtained in 
1964 \cite{Be64}. He proved that any {\it deterministic local
hidden--variable theory} is incompatible with some statistical
prediction of quantum mechanics. This is the content of Bell's theorem
in its original form, which has been then generalized \cite{Be71} 
to include {\it non--deterministic theories}. 
EPR's paradox was interpreted as the need for the introduction
of additional variables, in order to restore {\it completeness},
{\it relativistic causality} (namely {\it locality}) and {\it realism} in the theory
(the point of view of realism asserts that quantum systems have intrinsic and
well defined properties even when they are not subject to measurements).
In line with this requirement, Bell and other 
authors \cite{CHSH69,Wi70,CH74,Cl78} derived different inequalities suitable for
testing what has been called {\it local realism}. 

Once established the particularity of Bell's local
realism in connection with the predictions
of quantum mechanics, different experiments have been designed 
and carried out to test these theories. The oldest ones \cite{Cl78,As82} measured the
linear polarization correlations of photon pairs created in radiative
atomic cascade reactions or in electron--positron annihilations, 
whereas, more recently, parametric
down--conversion photon sources have been employed \cite{Ze99,Ti98,We98}. 
Essentially all the experiments performed until now (in optics and atomic physics)
have proved that the class of theories governed by Bell's 
theorem are unphysical: they showed the violation of Bell's inequalities
and were in good agreement with the statistical predictions of quantum mechanics.
Actually, to be precise, because of apparata non--idealities
and other technical problems,
supplementary assumptions are needed in the interpretation of the experiments,
and, consequently, no test employed to refute local realism
has been completely loophole free \cite{Ti98,Cl78,CaSa94}.
It is then important to continue performing experiments on 
correlation properties of many particle systems, possibly in new
sectors, especially in particle physics, where entangled $K^0\bar{K}^0$ and $B^0\bar{B}^0$
pairs are considerable examples. If future investigations will confirm
the violation of Bell's inequalities,
it is clear that, under the philosophy of realism,
the locality assumption would be incompatible with experimental evidence. 
Then, if this were the case, maintaining realism one should
consider as a real fact of Nature a non--local behaviour of quantum phenomena.
This fact is not in conflict with the theory of relativity. Actually, there is no
way to use quantum non--locality for faster--than--light communication: for a correlated
system of two separated entities, according to 
quantum mechanics, the result of a measurement on
a subsystem is always independent of the experimental setting used to measure
the other subsystem.

In this paper we discuss the predictions of
local realistic schemes for a pair of correlated neutral kaons created in the decay 
of the $\phi$--meson. The two--neutral--kaon system is the most interesting example
of massive two--particle system that can be employed to discuss descriptions of
microscopic phenomena alternative to quantum mechanics (for a discussion concerning 
possible violations of quantum mechanics in the $K^0$--$\bar{K}^0$
system see ref.~\cite{Pe95}). Unlike photons, kaons are detectable with high
efficiency (by observing $K_S$ and $K_L$ decays or
$K^0$ and $\bar{K}^0$ strong interactions with the nucleons of absorbers).
Moreover, for $K^0\bar{K}^0$ pairs, which can be copiously produced at a high
luminosity $\Phi$-factory, additional assumptions regarding detection
not implicit in local realism
(always implemented in the interpretation of experiments with
photon pairs \cite{Cl78}) are not necessary to derive Bell's inequalities suitable
for experimental tests of local realism \cite{Se90}. Finally, the two--kaon 
system offers the possibility
for tests on unexplored time and energy scales.
A correlation experiment discriminating between local realism 
and quantum mechanics could be performed at the Frascati $\Phi$-factory 
in the future \cite{Frasc}. Indeed, being designed to measure direct $CP$ violation
in the $K^0$--$\bar{K}^0$ system, such a factory employs high precision
detectors. Unlike the other papers in the literature 
\cite{Gh91,Eb93,DD95,Uc97,BF98,Br99,Se97,Gi00} which treated the 
two--kaon correlated system within local realistic models, 
we shall discuss tests of local realism both with
and without the use of Bell's inequalities.

The work is organized as follows. In section~\ref{localrealism} we introduce, 
starting from the original EPR's program, the point of view of
local realism for the two--kaon system. The quantum--mechanical expectation
values relevant for the evolution of the system are briefly summarized
in section~\ref{qmp}. Section~\ref{lrp} is devoted to the presentation of the
local realistic scheme we use to describe the observable behaviour of the
pair: the philosophy of realism is implemented in our discussion by
means of the most general hidden--variable interpretation of the two--kaon evolution. 
Then, in section~\ref{comp} we study the compatibility among the local realistic
expectation values and the statistical predictions of quantum mechanics. 
In agreement with Bell's theorem, we show how any local hidden--variable
theory for the two--kaon entangled state
is incompatible with certain predictions of quantum mechanics.
In section \ref{imposs} the difficulties of testing local realism 
for the $K^0$--$\bar{K}^0$ state
by employing Bell--type inequalities are discussed. We show that,
contrary to what is generally believed in the literature,
a Bell--type test at a $\Phi$--factory is possible.
Our conclusion are given in section~\ref{concl}.

\section{From EPR's argument to Local Realism}
\label{localrealism}
The starting point of EPR's argumentation was the following
condition for a \underline{complete theory}:
{\it every element of physical reality must have a counterpart in the physical theory}.
They defined the \underline{physical reality} by means of the following sufficient
criterion: {\it if, without in any way disturbing a system, we can predict with certainty
(i.e., with probability equal to unity) the value of a physical quantity, then there exists
an element of physical reality corresponding to this quantity}. 
In addition, for a system made of two correlated, spatially separated 
and non--interacting entities,
EPR introduced the following \underline{locality} assumption:
{\it since at the time of measurement the two systems no longer interact, no real
change can take place in the second system in consequence of anything that may
be done to the first system}. 

EPR assumed that the physical world is analyzable in terms of
distinct and separately existing elements of reality, which are represented,
in the supposed complete theory, by well defined mathematical entities. 
The previous criterion of reality supports the anthropocentric
point of view nowadays called {\it realism}: it asserts that quantum systems have
intrinsic and well defined properties even when they are not subject to
measurements. Under this philosophy, the existence of quantum world is
(as in classical physics) objective:
thus, any measurement performed on a quantum system must produce a result with a 
definite and predetermined value.

To exemplify EPR's argumentation, consider the case of a particle with
total angular momentum zero which decays, at rest, into two spin $1/2$ particles,
1 and 2, with zero relative orbital angular momentum, 
which fly apart with opposite momenta. After a certain time 
(when the particles are separated by a macroscopic distance) 
suppose they do not interact any 
more (this situation corresponds to the EPR--Bohm's gedanken experiment 
\cite{Bm51,Bm57}). 
At this time, the normalized spin wave function of the global system, 
which does not depend (because of the spherical symmetry of the singlet state) 
on the quantization direction of the spin, is:
\beq
\label{spin0}
|S=0,S_z=0\rangle=\frac{1}{\sqrt{2}}\left[|+\rangle_1|-\rangle_2
-|-\rangle_1|+\rangle_2\right] .
\eeq
For particles 1 and 2, $|+\rangle$ and $|-\rangle$ represent spin--up and spin--down states,
respectively, along a direction chosen as $z$--axis. Because of the entangled nature
of this wave function, the two particles
do not have definite values of the spin component along any direction. The superposition
of two product states (\ref{spin0}) produces then non--factorizable joint probabilities.
The paradoxical behaviour of correlated and non--interacting
systems originates from the fact that
the wave function of the global system is not a tensorial product
of superpositions of states of the component systems.

When 1 and 2 do not interact any more,
a measurement of the spin component of one particle produces a given outcome 
[which is not predetermined by the quantum state (\ref{spin0})] and
forces, {\it immediately}, the spin of the other particle along the opposite direction;
notice that this is independent of whether or not any measurement is
then performed on the other particle. For instance, if the result of a measurement along 
the $z$--axis finds particle 1 in the spin--up state, we conclude
that at the same time particle 2 (which is supposed not to interact with particle 1 nor with
the measuring device) has spin--down along $z$; the wave packet reduction has led
to the disentanglement of the superposition (\ref{spin0}):
\beq
|S=0,S_z=0\rangle\to |+\rangle_1|-\rangle_2 ,
\eeq
and the total angular momentum of the pair is indefinite after the measurement.
The instantaneous response (due to the collapse
of the wave function) of the particle which is not observed is what Einstein
called {\it spooky action--at--a--distance}.

The two particles of Bohm's gedanken experiment
are perfectly correlated, and, following EPR, the spin component
of particle 2 is an element of physical reality, since it is predicted 
{\it with certainty} and {\it without in any way disturbing} particle 2. Moreover, 
in order to fulfil the locality assumption (no action--at--a--distance),
EPR assumed that such an element of reality existed independently of any measurement
performed on particle 1. Following EPR's argumentation, 
the interpretation of the above experiment by means of quantum mechanics
lead to a difficulty. In fact, if we had performed a measurement of the spin component of 
particle 1 along another direction, say along the $x$-axis, this would have defined 
the $x$ component of the spin of particle 2 as another element of reality,
again independent of measurement. Obviously, this is also
valid for any spin component, then it should be possible, in the supposed complete
theory, to assign different spin wave functions
to the same physical reality. Therefore, one arrives at the conclusion that two or more physical
quantities, which correspond to non--commuting quantum operators, can have simultaneous reality.
However, in quantum mechanics two observables corresponding to
non-commuting operators cannot have simultaneous reality. 
Therefore, there exist elements of physical reality for which quantum 
mechanics has no counterpart, and, according to EPR's completeness definition, 
quantum theory cannot give a complete description of reality.

Actually, one could object, with Bohr \cite{Bo35}, that
in connection with a correlated system of non--interacting
subsystems [described, in quantum mechanics, by eq.~(\ref{spin0})], EPR's reality criterion
reveals the following weak point: it is not correct to assert that the measurement on
subsystem 1 does not disturb system 2; in fact, in quantum mechanics
the measurement do separate systems
1 and 2, which are not separated entities before the reduction of the wave
packet. It is the measurement on system 1 that fixes 
(in a way that, however, does not depend only on the experimental
setting one uses but contains an element of randomness) the quantum state
(before undetermined) of system 2. Any measurement on system 1 is therefore a
measurement on the entire system 1+2. 
Moreover, in quantum mechanics
two or more physical quantities can be considered as simultaneous elements
of reality only when they can be simultaneously measured. Then, from the point of view
of orthodox quantum mechanics, EPR's argumentation ceases to be a paradox:
EPR's proof of incompleteness is mathematically correct but is founded on premises which are
inapplicable to microphenomena. 

However, one has to remind that, since in quantum mechanics the
{\it elements of reality} of quantum systems are our {\it knowings} (and not elements 
concerning the actual behaviour of matter),
this interpretation only provides an {\it incomplete} description of the {\it dynamics} of
quantum world, because each knowing originates a collapse of the wave function;
this process affects the future behaviour of the system and
randomly selects among different and alternative possibilities, whose
only known characteristic is the statistical distribution. Thus, 
in quantum mechanics the reality
of two non--commuting observables, which cannot be defined simultaneously,
depends on the measurement one performs. In this way, reality is in part
created by the observer. 

EPR's paradox was interpreted as the need for the introduction,
in quantum mechanics, of additional variables, in order to restore 
{\it completeness}, {\it causality} and {\it realism}.
Then, Bell and other authors developed different 
inequalities suitable for testing what has been called
\underline{local realism}. 

\subsection{Local Realism for the two--neutral--kaon system}
\label{localrealismkappa}
Now we come to the entangled system of two neutral kaons. In this paper we neglect the effects
of $CP$ violation. Then, the $CP$ eigenstates are identified with the short and
long living kaons (mass eigenstates): $|K_+ \rangle \equiv |K_S \rangle$ ($CP=+1$),
$|K_- \rangle \equiv |K_L \rangle$ ($CP=-1$). In this approximation the strong
interaction eigenstates $|K^0 \rangle$ and $|\bar{K}^0 \rangle$ are given by:
\begin{eqnarray}
\label{s-eig}
|K^0 \rangle& = & \frac{1}{\sqrt 2}\left[|K_S\rangle + |K_L\rangle\right] , \\   
|\bar{K}^0 \rangle& = & \frac{1}{\sqrt 2}\left[|K_S\rangle - |K_L\rangle\right] .
\nonumber
\end{eqnarray}
The time evolution of the mass (weak interaction) eigenstates is:
\beq
\label{time}
|K_{S,L}(\tau) \rangle=e^{-i\lambda_{S,L}\tau}|K_{S,L}\rangle ,
\eeq
where $|K_{S,L}\rangle\equiv |K_{S,L}(0)\rangle$,
$\tau= t \sqrt{1-v^2}$ is the kaon proper time 
[$t$ ($v$) being the time (kaon velocity) measured in the laboratory frame] and:
\beq
\lambda_{S,L}=m_{S,L}-\frac{i}{2}\Gamma_{S,L} ,
\eeq
$m_{S,L}$ denoting the $K_S$ and $K_L$ masses and $\Gamma_{S,L}$ the
corresponding decay widths: $\Gamma_{S,L}\equiv 1/{\tau_{S,L}}$
(we use natural units: $\hbar =c=1$).

Consider now the strong decay of the $\phi(1020)$--meson, whose relevant quantum numbers are
$J^{PC}=1^{--}$, into $K^0\bar{K}^0$ [$BR(\phi\to K^0\bar{K}^0)\simeq 34.1\%$].
With good approximation the process is non--relativistic:
in the center of mass system, the kaons correspond to a Lorentzian 
factor $\gamma \simeq 1.02$. Just after the decay, at proper time $\tau=0$,
the quantum--mechanical state is given by the following superposition:
\begin{eqnarray}
\label{kkk}
|\phi(0)\rangle & = & \frac{1}{\sqrt 2}\left[  
|K^0\rangle_1 |\bar{K}^0\rangle_2 - |\bar{K}^0\rangle_1 |K^0\rangle_2\right] \\
& = & \frac{1}{\sqrt 2}\left[
|K_L\rangle_1 |K_S\rangle_2 - |K_S\rangle_1 |K_L\rangle_2\right] , \nonumber
\end{eqnarray}
written in both bases we have introduced. Since the kaon is a spinless particle
and the $\phi$ has spin 1, angular momentum conservation requires
the kaons to be emitted in a spatially antisymmetric state. The state is also
antisymmetric under charge conjugation. The second equality in (\ref{kkk}) is only approximated 
when one includes the (small) effects of $CP$ violation. Moreover, in the above equation, $1$ and $2$
denote the directions of motion of the two kaons. From eqs.~(\ref{s-eig}) and
(\ref{time}) the time evolution of state (\ref{kkk}) is obtained in the
following form:
\begin{eqnarray}
\label{qm2k}
|\phi(\tau_1,\tau_2)\rangle & = & \frac{1}{\sqrt 2}\left\{
e^{-i(\lambda_L\tau_1+\lambda_S\tau_2)}|K_L\rangle_1|K_S\rangle_2
-e^{-i(\lambda_S\tau_1+\lambda_L\tau_2)}|K_S\rangle_1|K_L\rangle_2\right\} \\ \nonumber
& = & \frac{1}{2\sqrt 2}\left\{
\left[e^{-i(\lambda_L\tau_1+\lambda_S\tau_2)}+
e^{-i(\lambda_S\tau_1+\lambda_L\tau_2)}\right]
\left[|K^0\rangle_1 |\bar{K}^0\rangle_2 - |\bar{K}^0\rangle_1 |K^0\rangle_2\right]\right. \\
\nonumber
& & \left. +\left[e^{-i(\lambda_L\tau_1+\lambda_S\tau_2)}-
e^{-i(\lambda_S\tau_1+\lambda_L\tau_2)}\right]
\left[|K^0\rangle_1 |K^0\rangle_2 - |\bar{K}^0\rangle_1 |\bar{K}^0\rangle_2\right]
\right\} .
\end{eqnarray}

In the following we introduce, within local realism,
the elements of physical reality for the two--kaon system.
Before doing this, it is important to remind once again that 
within the philosophy of realism, quantum systems have
intrinsic and well defined properties, even when they are not subject to
measurements. The existence of quantum world is then 
(like in classical physics) objective and independent of our observations.
As a consequence, any measurement performed on a quantum system produces a
result with a definite and predetermined value. We shall assume locality by requiring that
physical phenomena in a space--time region
cannot be affected by what occurs in all space--time regions which are
space--like separated from the first one. This means that
when the two kaons are space--like separated, the elements of reality
belonging to one kaon cannot be created nor
influenced by a measurement made on the
other kaon. This amounts to express relativistic causality, which
prevents any {\it action--at--a--distance}. Implicit in our
description is also the 
inexistence, in any reference frame, of influences acting backward in time:
a measurement performed on one kaon cannot influence the elements of reality
possessed by this kaon for times preceding the measurement.

Quantum mechanics predicts (and we know it is a well tested property)
a perfect anti--correlation in strangeness and $CP$ values
when both kaons are considered at the same time [see eq.~(\ref{qm2k})].
If an experimenter observes, say along direction 1, a $K^0$ ($K_L$), at the same time
$\tau_1$, along direction 2, because of the instantaneous collapse of the two--kaon wave function, 
one can predict the presence of a $\bar{K}^0$ ($K_S$). Thus, at time $\tau_1$ to the kaon 
moving along direction 2 we assign an element of reality
(since, following EPR's reality criterion the value of the corresponding physical quantity
is predicted {\it with certainty} and {\it without in any way disturbing}
the system), the value $-1$ ($+1$) of strangeness ($CP$). 
The same discussion is valid when the state observed
along direction 1 is $\bar{K}^0$ (or $K_S$) as well as when one exchanges the kaon
directions: $1\leftrightarrow2$. For times $\tau_2$ successive the observation at time $\tau_1$
along direction 1 of a $K_L$ ($K_S$), a $CP$
measurement on the other kaon will give with certainty the same result
$CP=+1$ ($CP=-1$) one expects at time $\tau_1$. This expresses $CP$ conservation.
Obviously, because of the instability of the $K_L$ and $K_S$
components, along direction 2 the experimenter could observe 
either $CP=+1$ or $CP=-1$ decay products at time $\tau_2$, but what is important 
in the present discussion is
that for any pair of times $(\tau_1,\tau_2)$ these exists perfect anti--correlation on $CP$. 
In the case in which both kaons are undecayed, when the kaon detected at time $\tau_1$
is $K^0$ ($\bar{K}^0$), at times $\tau_2>\tau_1$ along direction 2 quantum
mechanics predicts the possibility to observe a 
$\bar{K}^0$ ($K^0$) as well as a $K^0$ ($\bar{K}^0$):
since strangeness is not conserved during the evolution
of the system (governed by the weak interaction), perfect anti--correlation 
on strangeness only exists
when both particles are considered at the same time. 

Following EPR's argument, 
in the local realistic approach one then associates to both kaons of the pair,
at any time, two elements of reality, which are not created by
measurements eventually performed on the partner when the
particles are space--like separated (locality): one determines the 
kaon $CP$ value, the other one supplies the kaon strangeness $S$.
They are both well defined also when the meson is not observed (realism)
and can take two values, $\pm 1$, which appear at random with the same 
frequency in a statistical ensemble of kaons. Because of the strangeness
non--conservation, a particular value of the element of reality $S$ is
defined instantaneously (in fact, instantaneous oscillations between $S=\pm 1$
and $S=\mp 1$ occur), but what is important in the realistic approach is that 
$S$ has objective and well defined existence at any instant time. For a pair, 
the instantaneous and simultaneous $|\Delta S|=2$ oscillations are compatible with locality
only if one introduces a hidden--variable interpretation of the pair evolution
which predetermines the times of the strangeness jumps. 

In conclusion, neglecting $CP$ violation, within local realism a kaon is 
characterized by two different elements of physical reality,
which can both take two values with equal frequency; 
thus, four different single kaon states can
appear just after the $\phi$ decay, with the same
frequency (25\%). They are quoted in table \ref{loc-rea}. 
\begin{table}[t]
\begin{center}
\caption{Kaon realistic states.}
\label{loc-rea}
\begin{tabular}{c|c c}
\mc {1}{c|}{State} &
\mc {1}{c}{Strangeness} &
\mc {1}{c}{CP} \\ \hline
$K_1\equiv K^0_{S}$      & $+1$& $+1$  \\
$K_2\equiv \bar{K}^0_S$  & $-1$& $+1$  \\
$K_3\equiv K^0_L$        & $+1$& $-1$  \\
$K_4\equiv \bar{K}^0_L $ & $-1$& $-1$  \\
\end{tabular}
\end{center}
\end{table}
It is clear that this classification is incompatible with
quantum mechanics: in fact, under local realism a kaon has, simultaneously,
defined values of strangeness and $CP$, whereas in quantum mechanics these
quantities are described by non--commuting operators, then they cannot 
be measured simultaneously.

\section{Quantum--mechanical expectation values}
\label{qmp}

By introducing the shorthand notation:
\beq
E_{S,L}(\tau)=e^{-\Gamma_{S,L}\tau} ,
\eeq
and the mass difference:
\beq
\Delta m=m_L-m_S ,
\eeq
from eq.~(\ref{qm2k}) the quantum--mechanical (QM) probability 
$P_{QM}[K^0(\tau_1),\bar{K}^0(\tau_2)]\equiv |_1\langle K^0|_2\langle \bar{K}^0|
\phi(\tau_1,\tau_2)\rangle|^2$ ($P_{QM}[\bar{K}^0(\tau_1),K^0(\tau_2)]\equiv
|_1\langle \bar{K}^0|_2\langle K^0|\phi(\tau_1,\tau_2)\rangle|^2$)
that a measurement detects a $K^0$ ($\bar{K}^0$) at time 
$\tau_1$ along direction $1$ and a $\bar{K}^0$ ($K^0$) at time
$\tau_2$ along direction $2$ is:
\begin{eqnarray}
\label{kkb}
& &P_{QM}[K^0(\tau_1),\bar{K}^0(\tau_2)]=P_{QM}[\bar{K}^0(\tau_1),K^0(\tau_2)] \\
& & \nonumber \\
& &= \frac{1}{8}\left[E_L(\tau_1)E_S(\tau_2)+E_S(\tau_1)E_L(\tau_2)+
2\sqrt{E_L(\tau_1+\tau_2)E_S(\tau_1+\tau_2)}{\rm cos}\,\Delta m(\tau_2-\tau_1)\right] .
\nonumber
\end{eqnarray}
The other probabilities relevant for our discussion are the following ones:
\begin{eqnarray}
\label{kk}
& &P_{QM}[K^0(\tau_1),K^0(\tau_2)]=P_{QM}[\bar{K}^0(\tau_1),\bar{K}^0(\tau_2)] \\
& & \nonumber \\
& &= \frac{1}{8}\left[E_L(\tau_1)E_S(\tau_2)+E_S(\tau_1)E_L(\tau_2)-
2\sqrt{E_L(\tau_1+\tau_2)E_S(\tau_1+\tau_2)}{\rm cos}\,\Delta m(\tau_2-\tau_1)\right] ,
\nonumber
\end{eqnarray}
\begin{eqnarray}
\label{ls}
P_{QM}[K_L(\tau_1),K_S(\tau_2)]&=&\frac{1}{2}E_L(\tau_1)E_S(\tau_2) , \\
\label{sl}
P_{QM}[K_S(\tau_1),K_L(\tau_2)]&=&\frac{1}{2}E_S(\tau_1)E_L(\tau_2) , \\
\label{ss}
P_{QM}[K_S(\tau_1),K_S(\tau_2)]&=&P_{QM}[K_L(\tau_1),K_L(\tau_2)]=0 , \\
\label{skappa}
P_{QM}[K_S(\tau_1),K^0(\tau_2)]&=&P_{QM}[K_S(\tau_1),\bar{K}^0(\tau_2)]
=P_{QM}[K^0(\tau_1),K_L(\tau_2)] \\ 
&=&P_{QM}[\bar{K}^0(\tau_1),K_L(\tau_2)]=\frac{1}{4}E_S(\tau_1)E_L(\tau_2) , \nonumber \\
\label{lkappa}
P_{QM}[K_L(\tau_1),K^0(\tau_2)]&=&P_{QM}[K_L(\tau_1),\bar{K}^0(\tau_2)]
=P_{QM}[K^0(\tau_1),K_S(\tau_2)] \\
&=&P_{QM}[\bar{K}^0(\tau_1),K_S(\tau_2)]=\frac{1}{4}E_L(\tau_1)E_S(\tau_2) ,  \nonumber
\end{eqnarray}
the fourth equation expressing $CP$ conservation.

In the particular case of $\tau_1=\tau_2\equiv \tau$:
\begin{eqnarray}
P_{QM}[K^0(\tau),\bar{K}^0(\tau)]&=&P_{QM}[\bar{K}^0(\tau),K^0(\tau)]
=\frac{1}{2} E_L(\tau)E_S(\tau) , \\
\label{anti-corr}
P_{QM}[K^0(\tau),K^0(\tau)]&=&P_{QM}[\bar{K}^0(\tau),\bar{K}^0(\tau)]=0 . 
\end{eqnarray}
These relations, together with eq.~(\ref{ss}),
show the perfect anti--correlation of the quantum-mechanical state
(\ref{qm2k}) concerning strangeness and $CP$.

Starting from probabilities (\ref{kkb}) and (\ref{kk}) it is useful to introduce
a time--dependent asymmetry parameter, defined by the following 
relation for a generic theory:
\beq
\label{asimm}
A(\tau_1,\tau_2)\equiv \frac{P[K^0(\tau_1),\bar{K}^0(\tau_2)]+
P[\bar{K}^0(\tau_1),K^0(\tau_2)]-
P[K^0(\tau_1),K^0(\tau_2)]-P[\bar{K}^0(\tau_1),\bar{K}^0(\tau_2)]}
{P[K^0(\tau_1),\bar{K}^0(\tau_2)]+P[\bar{K}^0(\tau_1),K^0(\tau_2)]
+P[K^0(\tau_1),K^0(\tau_2)]+P[\bar{K}^0(\tau_1),\bar{K}^0(\tau_2)]} .
\eeq
The quantum--mechanical expression of this quantity
is a function of $\tau_2-\tau_1$ only:
\beq
\label{qmasymm}
A_{QM}(\tau_1,\tau_2)=2\frac{\sqrt{E_L(\tau_2-\tau_1)E_S(\tau_2-\tau_1)}}
{E_L(\tau_2-\tau_1)+E_S(\tau_2-\tau_1)}{\rm cos}\, \Delta m (\tau_2-\tau_1) ,
\eeq
and measures the interference term appearing in like--strangeness ($K^0K^0$ or $\bar{K}^0\bar{K}^0$)
and unlike--strangeness ($K^0\bar{K}^0$ or $\bar{K}^0K^0$) events.

\section{Local realistic expectation values}
\label{lrp}

In this section we discuss the widest class of 
local hidden--variable models for the two--kaon state
and their predictions for the observables
provided, in quantum mechanics, by eqs.~(\ref{kkb})--(\ref{lkappa}). 
Following the derivation of ref.~\cite{Se97}, we start considering 
how the quantum--mechanical expectation values for the single kaon evolution
can be reproduced by a realistic approach. 
Then, we extend the description to the interesting case of an entangled kaon pair.

\subsection{Evolution of a single kaon}
\label{singlek}
In the realistic approach one introduces the four kaonic
states of table~\ref{loc-rea}. $K_1$ is a state with defined
strangeness ($+1$) and $CP$ ($+1$), and the same is true for the other states.

Introduce the notation:
\beq
p_{ij}(\tau|0)\equiv p[K_j(0)\to K_i(\tau)] ,
\eeq
for the {\it conditional} probability that a state $K_i$ is present at time $\tau$
if the original state at time $\tau=0$ was $K_j$. It is immediate
to write down the time $\tau=0$ probabilities; they are:
\beq
\label{timezero}
p_{11}(0|0)=p_{22}(0|0)=p_{33}(0|0)=p_{44}(0|0)=1 , \hspace{0.3cm} p_{ij}(0|0)=0 
\hspace{0.2cm} {\rm for} \hspace{0.2cm} i\neq j .
\eeq
When the evolution of the four states is considered, $CP$ conservation requires
that, for all times:
\begin{eqnarray}
\label{cp-cons}
p_{13}(\tau|0)&=&p_{14}(\tau|0)=p_{23}(\tau|0)=p_{24}(\tau|0)=p_{31}(\tau|0) \\
&=&p_{32}(\tau|0)=p_{41}(\tau|0)=p_{42}(\tau|0)\equiv 0 . \nonumber
\end{eqnarray}
In quantum mechanics, assuming $CP$ conservation, the mass eigenstates $|K_L \rangle$ 
and $|K_S \rangle$ are perfectly orthogonal to each other, then:
\beq
\langle K_L(0)|K_S(\tau)\rangle = \langle K_S(0)|K_L(\tau)\rangle =0 .
\eeq
During the time evolution, strangeness jumps between $S=+1$ ($-1$) and $S=-1$ ($+1$) states occur.
Thus, only transitions $K_1\leftrightarrow K_2$ and $K_3\leftrightarrow K_4$ are permitted,
and eq.~(\ref{cp-cons}) is valid. 

In order to fix the time evolution of the four states
of table~\ref{loc-rea} we have to determine 8 probabilities $p_{ij}(\tau|0)$.
As we are going to show, it is possible to fix these quantities and reproduce all the 
quantum--mechanical predictions relevant for the single kaon
propagation \cite{Se97}. 

From quantum mechanics [eqs.~(\ref{s-eig}), (\ref{time})] one obtains:
\begin{eqnarray}
|\langle K^0(0)|K_S(\tau)\rangle|^2&=&|\langle \bar{K}^0(0)|K_S(\tau)\rangle|^2=
|\langle K_S(0)|K^0(\tau)\rangle|^2 \\
&=&|\langle K_S(0)|\bar{K}^0(\tau)\rangle|^2=\frac{1}{2}E_S(\tau) , \nonumber
\end{eqnarray}
where the different terms have obvious significance.
This restrictions correspond to require the following equalities, 
that we write in the same order as before, among the realistic probabilities:
\begin{eqnarray}
\frac{1}{2}[p_{11}(\tau|0)+p_{12}(\tau|0)]&=&\frac{1}{2}[p_{21}(\tau|0)+p_{22}(\tau|0)]=
\frac{1}{2}[p_{11}(\tau|0)+p_{21}(\tau|0)] \\
&=&\frac{1}{2}[p_{12}(\tau|0)+p_{22}(\tau|0)]= \frac{1}{2}E_S(\tau) , \nonumber
\end{eqnarray}
which correspond to fix:
\begin{eqnarray}
\label{21}
p_{21}(\tau|0)&=&p_{12}(\tau|0) , \\
\label{22}
p_{22}(\tau|0)&=&p_{11}(\tau|0) , \\
\label{prima}
p_{11}(\tau|0)+p_{12}(\tau|0)&=&E_S(\tau) ,
\end{eqnarray}
the first two equalities being compatible with time--reversal invariance, which follows from 
$CPT$ theorem, having adopted $CP$ conservation. 
In the same way, the equalities:
\begin{eqnarray}
|\langle K^0(0)|K_L(\tau)\rangle|^2&=&|\langle \bar{K}^0(0)|K_L(\tau)\rangle|^2=
|\langle K_L(0)|K^0(\tau)\rangle|^2 \\
&=&|\langle K_L(0)|\bar{K}^0(\tau)\rangle|^2= \frac{1}{2}E_L(\tau) , \nonumber
\end{eqnarray}
require:
\begin{eqnarray}
\label{43}
p_{43}(\tau|0)&=&p_{34}(\tau|0) , \\
\label{44}
p_{44}(\tau|0)&=&p_{33}(\tau|0) , \\
\label{seconda}
p_{33}(\tau|0)+p_{34}(\tau|0)&=&E_L(\tau) .
\end{eqnarray}
At this point, two of the 8 $p_{ij}$'s are independent.
However, other constraints come from quantum mechanics.
In fact, one can write:
\begin{eqnarray}
|\langle K^0(0)|K^0(\tau)\rangle|^2&=&\frac{1}{4}\left[E_L(\tau)+E_S(\tau)
+2\sqrt{E_L(\tau)E_S(\tau)}{\rm cos}\,\Delta m \tau\right] \\
& = & \frac{1}{2}[p_{11}(\tau|0)+p_{33}(\tau|0)] , \nonumber
\end{eqnarray}
where the first (second) equality follows from quantum mechanics (realism),
and, analogously:
\begin{eqnarray}
|\langle \bar{K}^0(0)|K^0(\tau)\rangle|^2&=&\frac{1}{4}\left[E_L(\tau)+E_S(\tau)
-2\sqrt{E_L(\tau)E_S(\tau)}{\rm cos}\,\Delta m \tau\right] \\
& = & \frac{1}{2}[p_{12}(\tau|0)+p_{34}(\tau|0)] , \nonumber
\end{eqnarray}
where, in the last equality, we have taken into account of eqs.~(\ref{21}) and
(\ref{43}). The other equations one gets for the quantities
$|\langle K^0(0)|\bar{K}^0(\tau)\rangle|^2$,
$|\langle \bar{K}^0(0)|\bar{K}^0(\tau)\rangle|^2$,
$|\langle K_S(0)|K_S(\tau)\rangle|^2$ and 
$|\langle K_L(0)|K_L(\tau)\rangle|^2$ do not supply new constraints but 
are compatible with the conditions written above. Thus, assuming
eqs.~(\ref{21}), (\ref{22}), (\ref{43}) and (\ref{44}),
among $p_{11}$, $p_{12}$, $p_{33}$ and $p_{34}$ we have the system of equations:
\beq
\left\{\begin{array}{l}
\label{1112}
p_{11}(\tau|0)+p_{12}(\tau|0)=E_S(\tau)  \\
p_{33}(\tau|0)+p_{34}(\tau|0)=E_L(\tau)  \\
p_{11}(\tau|0)+p_{33}(\tau|0)=[E_L(\tau)+E_S(\tau)]Q_+(\tau)  \\
p_{12}(\tau|0)+p_{34}(\tau|0)=[E_L(\tau)+E_S(\tau)]Q_-(\tau) ,
\end{array}\right. 
\eeq
where the shorthand notation:
\beq
Q_{\pm}(\tau)=\frac{1}{2}\left[1\pm 2\frac{\sqrt{E_L(\tau)E_S(\tau)}}
{E_L(\tau)+E_S(\tau)}{\rm cos}\,\Delta m \tau \right]
\eeq
has been employed. Since $Q_+(\tau)+Q_-(\tau)=1$, in eq.~(\ref{1112})
only three conditions out of four are independent. 
A symmetrical choice of $p_{11}$, $p_{12}$, $p_{33}$ and $p_{34}$
leads to the following realistic probability matrix \cite{Se97}:
\begin{eqnarray}
\label{matrix}
&&{\bf p}(\tau|0) \\
&&=\left(\begin{array}{c c c c}
E_S(\tau)Q_+(\tau) + \delta(\tau) & E_S(\tau)Q_-(\tau) - \delta(\tau) & 0 & 0 \\
E_S(\tau)Q_-(\tau) - \delta(\tau) & E_S(\tau)Q_+(\tau) + \delta(\tau) & 0 & 0 \\
0 & 0 & E_L(\tau)Q_+(\tau) - \delta(\tau) & E_L(\tau)Q_-(\tau) + \delta(\tau)  \\
0 & 0 & E_L(\tau)Q_-(\tau) + \delta(\tau) & E_L(\tau)Q_+(\tau) - \delta(\tau)
\end{array}\right), \nonumber
\end{eqnarray}
where the degree of freedom is given by the function $\delta(\tau)$.
The requirement that all the matrix elements are well defined
($0\leq p_{ij}(\tau|0)\leq 1$) can be satisfied if one
chooses properly the function $\delta(\tau)$.
A particular solution correspond to keep $\delta(\tau)\equiv 0$. Actually, 
as we shall see in section \ref{evalprob}, an identically vanishing 
$\delta(\tau)$ function is the only solution compatible with 
the local realistic evolution of a correlated pair of kaon.

\subsection{Evolution of a correlated kaon pair}   
\label{pair}
Now we come to the time evolution of a correlated 
and non--interacting $K^0\bar{K}^0$ pair emitted in the
decay of a $\phi$--meson. 

From quantum theory [eq.~(\ref{qm2k})] we know that for any time the joint
observation of the mesons finds them perfectly correlated.
At time $\tau =0$, immediately after the $\phi$ decay, in the realistic
picture there are four possible states for the kaon pair, each appearing 
with a probability equal to $1/4$: they are listed in table~\ref{corr0}.
\begin{table}[t]
\begin{center}
\caption{Realistic states for the kaon pair at initial time $\tau =0$.}
\label{corr0}   
\begin{tabular}{c c}
\mc {1}{c}{Direction 1} &
\mc {1}{c}{Direction 2} \\ \hline
$K_1\equiv K^0_{S}$ \hspace{0.4cm}       ($S=+1$, $CP=+1$)  &   
$K_4\equiv \bar{K}^0_{L}$ \hspace{0.4cm} ($S=-1$, $CP=-1$) \\
$K_2\equiv \bar{K}^0_S$ \hspace{0.4cm}   ($S=-1$, $CP=+1$)  & 
$K_3\equiv K^0_L$ \hspace{0.4cm}         ($S=+1$, $CP=-1$) \\
$K_3\equiv K^0_L$ \hspace{0.4cm}         ($S=+1$, $CP=-1$)  & 
$K_2\equiv \bar{K}^0_S$ \hspace{0.4cm}   ($S=-1$, $CP=+1$)  \\
$K_4\equiv \bar{K}^0_L$ \hspace{0.4cm}   ($S=-1$, $CP=-1$)  &
$K_1\equiv K^0_S$ \hspace{0.4cm}         ($S=+1$, $CP=+1$)   \\
\end{tabular}
\end{center}
\end{table}
Kaon $K_1$ is created together with a $K_4$: we assume,
as in quantum mechanics, since it is a well tested property, 
a perfect anti--correlation in strangeness and $CP$
when both kaons are considered at equal time. 
The other three initial states show, obviously, the same correlation property.

When the system evolves, the kaons fly apart from each other,
and at two generic times $\tau_1$ and $\tau_2$
(corresponding to opposite directions of propagation labeled 1 and 2, respectively)
the kaon pair is in one of the states reported in table~\ref{corrt}.
\begin{table}[t]
\begin{center}
\caption{Local realistic states for the kaon pair at times $\tau_2\geq \tau_1$.}
\label{corrt}
\begin{tabular}{l | c c}
\mc {1}{c |}{Probabilities}&
\mc {1}{c}{Direction 1 (Left) \hspace{0.2cm}Time $\tau_1$} &
\mc {1}{c}{Direction 2 (Right)\hspace{0.2cm}Time $\tau_2$} \\ \hline
$P_1(\tau_1,\tau_2;\lambda)$ & $K_1\equiv K^0_{S}$      &  $K_4\equiv \bar{K}^0_L$  \\
$P_2(\tau_1,\tau_2;\lambda)$ & $K_1\equiv K^0_{S}$      &  $CP=-1$ DP      \\  
$P_3(\tau_1,\tau_2;\lambda)$ & $CP=+1$ DP     &   $K_4\equiv \bar{K}^0_L$  \\
$P_4(\tau_1,\tau_2;\lambda)$ & $K_1\equiv K^0_{S}$      &  $K_3\equiv K^0_L$  \\

$P_5(\tau_1,\tau_2;\lambda)$ & $K_2\equiv \bar{K}^0_S$  &  $K_3\equiv K^0_L$        \\
$P_6(\tau_1,\tau_2;\lambda)$ & $K_2\equiv \bar{K}^0_S$  &  $CP=-1$ DP      \\
$P_7(\tau_1,\tau_2;\lambda)$ & $CP=+1$ DP     &   $K_3\equiv K^0_L$        \\  
$P_8(\tau_1,\tau_2;\lambda)$ & $K_2\equiv \bar{K}^0_S$  &  $K_4\equiv \bar{K}^0_L$   \\

$P_9(\tau_1,\tau_2;\lambda)$    & $K_3\equiv K^0_L$        &  $K_2\equiv \bar{K}^0_S$ \\
$P_{10}(\tau_1,\tau_2;\lambda)$ & $K_3\equiv K^0_L$        &  $CP=+1$ DP      \\
$P_{11}(\tau_1,\tau_2;\lambda)$ & $CP=-1$ DP     &   $K_2\equiv \bar{K}^0_S$  \\
$P_{12}(\tau_1,\tau_2;\lambda)$ & $K_3\equiv K^0_L$        &  $K_1\equiv K^0_S$ \\

$P_{13}(\tau_1,\tau_2;\lambda)$ & $K_4\equiv \bar{K}^0_L$  &  $K_1\equiv K^0_S$        \\
$P_{14}(\tau_1,\tau_2;\lambda)$ & $K_4\equiv \bar{K}^0_L$  &  $CP=+1$ DP      \\
$P_{15}(\tau_1,\tau_2;\lambda)$ & $CP=-1$ DP    &   $K_1\equiv K^0_S$  \\   
$P_{16}(\tau_1,\tau_2;\lambda)$ & $K_4\equiv \bar{K}^0_L$  &  $K_2\equiv \bar{K}^0_S$ \\
$P_{17}(\tau_1,\tau_2;\lambda)$ & $CP=+1$ DP & $CP=-1$ DP \\
$P_{18}(\tau_1,\tau_2;\lambda)$ & $CP=-1$ DP & $CP=+1$ DP \\
\end{tabular}
\end{center}
\end{table}
The first row refers to the state with a $K_1$ at time $\tau_1$ along direction
$1$ (which we define as left direction)
and a $K_4$ at time $\tau_2$ along direction $2$ (right direction;
we have in mind, here, the kaon pair propagation
in the center of mass system). Given the classification of the table, 
in our discussion we consider $\tau_2\geq\tau_1$: the isotropy of space
guarantees the invariance of the two--kaon states by exchanging the directions 1 and 2.
In the second row the state corresponds to a left going $K_1$ at time $\tau_1$ 
and $CP=-1$ decay products (DP) at time $\tau_2$ on the right. 
These decay products originate from the instability
of the $K_3$ and $K_4$ pure states, which are both long living kaons, namely
$CP=-1$ states (the corresponding physical processes are: 
$K_L\to 3\pi, \pi \mu \nu_{\mu}, \pi e \nu_e$). At time $\tau_1$ the state correlated
with a left going $K_1$ is necessarily either a $K_4$ or a state containing 
$CP=-1$ decay products, $K^{DP}_3$ or $K^{DP}_4$. 
Then, at time $\tau_2$ ($> \tau_1$) on the right we can have:
i) a $K_4$ (state in the first row), ii) $CP=-1$ decay products (state in
the second row) or iii) a $K_3$ (state in the fourth row). The former case refers to the
transition $K_4(\tau_1)\to K_4(\tau_2)$, the latter to
$K_4(\tau_1)\to K_3(\tau_2)$, both along direction 2. 
Occurrence ii) takes contributions
from the following transitions: $K^{DP}_3(\tau_1)\to K^{DP}_3(\tau_2)$,
$K^{DP}_4(\tau_1)\to K^{DP}_4(\tau_2)$,
$K_4(\tau_1)\to K^{DP}_4(\tau_2)$ and $K_4(\tau_1)\to K_3(\tau_1<\tau<\tau_2)
\to K^{DP}_3(\tau_2)$. The other states in table~\ref{corrt} have similar meaning,
the last two rows corresponding to the situation in which both left and right going kaons 
are decayed at times $\tau_1$ and $\tau_2$, respectively.

\subsubsection{Interpretation of the states with local hidden--variables} 
\label{lrstates} 
At this point it is important to stress that the states listed in 
table~\ref{corrt} are assumed to be well defined
for all times $\tau_1$ and $\tau_2$ with $\tau_1\leq \tau_2$: 
this is the main requirement of the 
realistic approach (analogous discussion is valid for states in tables~\ref{loc-rea}
and \ref{corr0}). For a given kaon pair we assume that
only one of the 18 possibilities of table~\ref{corrt} really occurs
for fixed $\tau_1$ and $\tau_2$. This means that we are making the
hypothesis (realism) that there exist additional variables, usually called
{\it hidden--variables} (with respect to orthodox quantum mechanics these
variables are hidden in the sense that they are uncontrollable), that provide a 
complete description of the pair, which is viewed as really existing
and with well defined properties independently of any observation. 
The state representing the meson pair for given times $(\tau_1,\tau_2)$
is completely defined by these hidden--variables:
they are supposed to determine in advance (say when
the two kaons are created) the future behaviour of the pair. Thus,
the times in correspondence of which the instantaneous $|\Delta S|=2$ jumps and the decay
occur for a given kaon are predetermined by its hidden--variables.
Under this hypotheses there is no problem concerning a possible causal influence
acting among the different entities of entangled systems when a measurement
takes place on one subsystem. 
However, the new variables, which we denote with the compact symbol $\lambda$, are 
unobservable because they are averaged out in the measuring processes, and  
unobservable are the states of table~\ref{corrt}. 
In principle, also the measuring apparata could be
described by means of hidden--variables, which influence the results of measurement.
Besides, hidden--variables associated to the kaon pair could show a non--deterministic
behaviour. 
It is important to stress that in the approach with hidden--variables, the probabilistic
character of quantum mechanics is viewed as a practical necessity
for treating problems at the observation level, but
(and this is a strong difference compared to the orthodox interpretation)
does not originate from the intrinsic
behaviour of microphenomena: the indetermination principle is supposed
to act only during the observation process.

The realistic probabilities listed in table \ref{corrt}:
\beq
P_i(\tau_1,\tau_2;\lambda)\equiv P_i(\tau_1,\tau_2|\lambda)\rho(\lambda) ,
\eeq
correspond to the situation in which a single meson pair, described by the value $\lambda$
of the hidden--variables, is considered. Once $\tau_1$ and $\tau_2$ are fixed, 
the {\it state} of the kaon pair $\lambda$ (we can think it is fixed at the time of the 
pair creation) can take values in the set $\{\lambda^{[\tau_1,\tau_2]}_i;\; i=1,..,18\}$
(however, we stress again, $\lambda$ is fixed when a single pair
is considered), and for a deterministic theory we have:
\beq
\label{hv-prob}
P_i\left(\tau_1,\tau_2;\lambda^{[\tau_1,\tau_2]}_j\right)=
\delta_{ij}\rho\left(\lambda^{[\tau_1,\tau_2]}_i\right) .
\eeq
In the previous relations, $\rho$ is the probability distribution of the 
kaon pair hidden--variables and $P_i(\tau_1,\tau_2|\lambda^{[\tau_1,\tau_2]}_j)$
($=\delta_{ij}$) is the probability of the $i$--th state of table 
\ref{corrt} {\it conditional} on the 
presence of a pair in the state $\lambda^{[\tau_1,\tau_2]}_j$. 
For a single meson pair, only one of the probabilities of table \ref{corrt} 
is different from zero at instants $(\tau_1,\tau_2)$
in a deterministic model:
if $\lambda\equiv \lambda^{[\tau_1,\tau_2]}_k$, the non--vanishing 
probability is the $k$--th of the table.
As far as different times $\tau'_1$ and $\tau'_2$ are considered, eq.~(\ref{hv-prob})
is valid for the same set of hidden--variables, but in general with
the variables appearing in a different permutation, 
$\{\lambda^{[\tau'_1,\tau'_2]}_i\}\equiv {\cal P}\{\lambda^{[\tau_1,\tau_2]}_j\}$
(indexes $i$ and $j$ always refer to the classification of table~\ref{corrt}). 
Therefore, in the model we are describing the pair can be created in 18 different 
realistic states, namely with 18 different values of the hidden--variables.
We stress again: the hidden--variable sets at different pair of times contain
the same objects, but in one of the 18! different orderings, 
and notation used in eq.~(\ref{hv-prob})
does not mean that the hidden--variables are time--dependent.
From eq.~(\ref{hv-prob}) it follows that in a model with deterministic 
kaon pair hidden--variables, the normalization of the local realistic probabilities 
corresponds to that of the hidden--variables:
\beq
\sum^{18}_{i=1} P_i\left(\tau_1,\tau_2;\lambda^{[\tau_1,\tau_2]}_i\right)
=\sum^{18}_{i=1}\rho\left(\lambda^{[\tau_1,\tau_2]}_i\right)=1 .
\eeq

In the realistic interpretation, the quantum state (\ref{qm2k}) corresponds to a 
statistical ensemble of meson pairs, which are
further specified by different values of the hidden--variables.
To exemplify,
let us consider such a (large) ensemble of identical kaon pairs specified by
different $\lambda$'s (which we suppose now to be continuous variables), 
whose distribution $\rho$ we assume to be independent of the {\it apparatus parameters}
$\tau_1$ and $\tau_2$, the kaons being emitted in a way which does not depend 
on the adjustable times $\tau_1$ and $\tau_2$ (we assume, here, no retroactive causality).
We can give a statistical characterization of this ensemble by means of the
set of observables (\ref{kkb})--(\ref{lkappa}). Considering, as an example, 
$P[K^0(\tau_1),\bar{K}^0(\tau_2)]$, within a general 
deterministic local hidden--variable interpretation:
\begin{eqnarray}
\label{hv}
P_{LR}[K^0(\tau_1),\bar{K}^0(\tau_2)]&\equiv &\int d\lambda \rho(\lambda)
P(K^0, \tau_1; \bar{K}^0, \tau_2|\lambda) \\
&=&\int d\lambda \rho(\lambda)
P^{\rm Left}(K^0,\tau_1|\lambda)P^{\rm Right}(\bar{K}^0,\tau_2|\lambda) , \nonumber
\end{eqnarray}
where $LR$ stands for local realism. The joint probability 
$P(K^0, \tau_1; \bar{K}^0, \tau_2|\lambda)$, which is conditional on the presence 
of the particular value $\lambda$ of the hidden--variables, has been assumed to be locally
explicable, then it appears in the factorized form in the last 
equality. The function $P^{\rm Left}(K^0,\tau_1|\lambda)$ 
[$P^{\rm Right}(\bar{K}^0,\tau_2|\lambda)$] 
is the conditional probability that, once fixed $\lambda$, 
the left (right) going kaon at time $\tau_1$ ($\tau_2$), which is fully
specified by $\lambda$, is $K^0$ ($\bar{K}^0$). As required by locality,
given $\lambda$ and the apparatus parameters $\tau_1$ and $\tau_2$,
$P^{\rm Left}$ and $P^{\rm Right}$ are independent. 
They only take two values:
\beq
\label{hv2}
P^{\rm Left}(K^0,\tau_1|\lambda)=\left\{\begin{array}{l}
1 , \hspace{0.25cm}{\rm when\; the\; state\; at\; time\;} \tau_1\; {\rm is\;} K^0 \\
0 , \hspace{0.25cm}{\rm when\; the\; state\; at\; time\;} \tau_1\; 
{\rm is\; not\;} K^0 \end{array}\right. .
\eeq
The knowledge (impossible, we emphasize again)
of the hidden--variables associated to an individual kaon pair 
emission would permit to determine the precise instants the $K_S$ and $K_L$
components decay, then eq.~(\ref{hv2}) follows.
The locality condition is motivated by the requirement of relativistic causality,
which prevents faster--than--light influences
between space--like separated events. In the present case, assuming there is no delay among
the times at which the experimenters choose to perform their observations and the real
kaon measurement times $\tau_1$ and $\tau_2$, the locality requirement is fulfilled when 
the two observation events are separated by a space--like interval [see eq.~(\ref{loc-exp})].
However, to be precise, as we shall explain in section \ref{comp}, a 
loophole that is impossible to
block exists and could permit, in principle and without requiring the existence of 
action--at--a--distance, an information to reach both the measuring devices for any choice of the
detection times $\tau_1$ and $\tau_2$. 

In the above we have restricted our argumentation to 
deterministic theories only, but it is possible to extend the same description
given by eq.~(\ref{hv}) to non--deterministic
(namely stochastic) theories as well as to deterministic theories in which additional
hidden--variables correspond to the measurement devices \cite{Cl78}. 
Let us make the hypothesis that also the experimental apparata are described in 
terms of hidden--variables, which influence the measurement outcomes. 
In this case, denoting with $\lambda'$ ($\lambda''$)
the hidden--variables specifying the behaviour of the apparatus measuring
on the left (right), in the new local hidden--variable
theory the expectation value of eq.~(\ref{hv}) is given by:
\begin{eqnarray}
\label{hv3}
P_{LR}[K^0(\tau_1),\bar{K}^0(\tau_2)]&\equiv &\int d\lambda\, d\lambda'\, d\lambda''
\rho(\lambda,\lambda',\lambda'') P(K^0, \tau_1; \bar{K}^0, \tau_2|
\lambda,\lambda',\lambda'') \\  
&=&\int d\lambda\, d\lambda'\, d\lambda'' \rho(\lambda)p(\lambda'|\lambda)p(\lambda''|\lambda)
P^{\rm Left}(K^0,\tau_1|\lambda,\lambda')
P^{\rm Right}(\bar{K}^0,\tau_2|\lambda,\lambda'') , \nonumber
\end{eqnarray}
where:
\beq
\label{rhorho}
\rho(\lambda,\lambda',\lambda'')\equiv p(\lambda',\lambda''|\lambda)\rho(\lambda)=
p(\lambda'|\lambda)p(\lambda''|\lambda)\rho(\lambda) .
\eeq
In the second equality of eq.~(\ref{hv3}), 
locality has been assumed for both kaon pair and apparata hidden--variables;
$p(\lambda'|\lambda)$ [$p(\lambda''|\lambda)$] is the conditional probability that,
when the kaon pair is specified by the variables $\lambda$, the 
device measuring the left (right) going kaon
is described by the variables $\lambda'$ ($\lambda''$). 
Since the distribution $\rho(\lambda)$ of the kaon pair hidden--variables is 
normalized to unity, the same occurs for
$p(\lambda'|\lambda)$ and $p(\lambda''|\lambda)$ for any $\lambda$.
It is than clear, by comparing eqs.~(\ref{hv}) and (\ref{hv3}), that:
\beq
\label{non-det}
P(K^0, \tau_1; \bar{K}^0, \tau_2|\lambda)=\int d\lambda'\, d\lambda''
p(\lambda',\lambda''|\lambda)
P(K^0,\tau_1;\bar{K}^0,\tau_2|\lambda,\lambda',\lambda'') , 
\eeq
and when one implements locality for the kaon pair 
and apparata hidden--variables, equality (\ref{hv2}) is replaced by:
\beq
\label{non-det2}
0\leq P^{\rm Left}(K^0,\tau_1|\lambda)\equiv\int d\lambda'\, p(\lambda'|\lambda)
P^{\rm Left}(K^0,\tau_1|\lambda,\lambda') \leq 1 .
\eeq
The difference compared to the deterministic case without apparata hidden--variables
is now clear, and in the new picture an equality like (\ref{hv2}) is valid for 
$P^{\rm Left}(K^0,\tau_1|\lambda,\lambda')$. It is important to stress
here that for the most general
non--deterministic local hidden--variable theory, elements of randomness
entering the probabilities $P^{\rm Left}(K^0,\tau_1|\lambda)$ and 
$P^{\rm Right}(\bar{K}^0,\tau_2|\lambda)$
could be related not only to apparata hidden--variables, but to other 
unknown mechanisms. Nevertheless, the above
discussion that have led to eqs.~(\ref{hv3})--(\ref{non-det2}) also applies for 
the most general non--deterministic local hidden--variable theory.
 
In our local realistic theory (table \ref{corrt})
the set of hidden--variables describing the kaon pair
forms a discrete set, and eq.~(\ref{hv}) 
reduces to:
\begin{eqnarray}
\label{ourhv}
P_{LR}[K^0(\tau_1),\bar{K}^0(\tau_2)]&=&\rho\left(\lambda^{[\tau_1,\tau_2]}_1\right)
P\left(K^0_S, \tau_1;\bar{K}^0_L,\tau_2|\lambda^{[\tau_1,\tau_2]}_1\right) \\
&&+ \rho\left(\lambda^{[\tau_1,\tau_2]}_9\right)
P\left(K^0_L, \tau_1;\bar{K}^0_S, \tau_2|\lambda^{[\tau_1,\tau_2]}_9\right) \nonumber \\
&\equiv &P_1\left(\tau_1,\tau_2;\lambda^{[\tau_1,\tau_2]}_1\right)+ 
P_9\left(\tau_1,\tau_2;\lambda^{[\tau_1,\tau_2]}_9\right) , \nonumber
\end{eqnarray}
where in the last line the notation of table \ref{corrt} has been introduced.
It is important to stress that, 
contrary to what occurs for quantum--mechanical probabilities, in the description 
with hidden--variables of
eqs.~(\ref{hv}) and (\ref{ourhv}) the transitions that lead to 
two--kaon states which contribute to 
$P_{LR}[K^0(\tau_1),\bar{K}^0(\tau_2)]$ do not interfere with one another.

Within the scheme of table~\ref{corrt} it is easy to obtain the predictions 
of local realism for measurements concerning an individual kaon. For instance, the 
probability to observe a left going $K^0$ at time $\tau_1$ is given by:
\beq
\label{ourhv2}
P^{\rm Left}_{LR}[K^0(\tau_1)]=\sum_{i=1,2,4,9,10,12}
P_i\left(\tau_1,\tau_2;\lambda^{[\tau_1,\tau_2]}_i\right) .
\eeq

The hidden--variable interpretation of the observables we have discussed in 
this section has also been assumed,
even if not explicitly declared, when, in section~\ref{singlek}, we treated the
propagation of a single kaon from the point of view of realism.

\subsubsection{Evaluation of the observables}
\label{evalprob}

Now we proceed discussing the range of variability of the 
meson pair observables compatible with the most general local realistic model.
We shall use the rules of classical probability theory. 

Start considering the probabilities of table~\ref{corrt}. In particular,
we concentrate on the state in the fourth row. 
At time $\tau_1$ the left going kaon is $K_1$, then, requiring $CP$ conservation, 
at time $\tau=0$ the initial state was either a $K_1$ or a $K_2$.
Since either a $K_1$ or a $K_2$ must be present as initial state for the
left going kaon (both with equal frequency $1/4$), from matrix~(\ref{matrix})
the probability that at time $\tau_1$ the state is $K_1$ equals to 
$[p_{11}(\tau_1|0)+p_{12}(\tau_1|0)]/4=E_S(\tau_1)/4$.
Correlated with this $K_1$, at the same time $\tau_1$ on the right
there is either a $K_4$ or $CP=-1$ decay products. Since the two--kaon state 
we are considering corresponds to a $K_3$ at time $\tau_2$, at time $\tau_1$ 
we must require the presence of a $K_4$: the probability that at this time
a $K_4$ is not decayed is $p_{43}(\tau_1|0)+p_{44}(\tau_1|0)=E_L(\tau_1)$.
Finally, from $\tau_1$ the $K_4$ must evolve into $K_3$ at time $\tau_2$. 
This transition occurs with (conditional) probability that we denote:
\beq
p_{34}(\tau_2|\tau_1)\equiv p[K_4(\tau_1)\to K_3(\tau_2)] .
\eeq
Then, probability $P_4$ of table~\ref{corrt} is given by:
\beq
\label{setprob0} 
P_4\left(\tau_1,\tau_2;\lambda^{[\tau_1,\tau_2]}_4\right)=
\frac{1}{4}E_S(\tau_1)E_L(\tau_1)p_{34}(\tau_2|\tau_1) .
\eeq
By using the same line of reasoning one obtains the other probabilities.
They have the following expressions:
\begin{eqnarray}
\label{setprob}
P_1\left(\tau_1,\tau_2;\lambda^{[\tau_1,\tau_2]}_1\right)&=&
\frac{1}{4}E_S(\tau_1)E_L(\tau_1)p_{44}(\tau_2|\tau_1) , \\ \nonumber
P_2\left(\tau_1,\tau_2;\lambda^{[\tau_1,\tau_2]}_2\right)&=&
P_6\left(\tau_1,\tau_2;\lambda^{[\tau_1,\tau_2]}_6\right)=\frac{1}{4}E_S(\tau_1)[1-E_L(\tau_2)] , \\ \nonumber
P_3\left(\tau_1,\tau_2;\lambda^{[\tau_1,\tau_2]}_3\right)&=&
\frac{1}{4}[1-E_S(\tau_1)]E_L(\tau_1)
[p_{43}(\tau_2|\tau_1)+p_{44}(\tau_2|\tau_1)] ,\\ \nonumber
P_5\left(\tau_1,\tau_2;\lambda^{[\tau_1,\tau_2]}_5\right)&=&
\frac{1}{4}E_S(\tau_1)E_L(\tau_1)p_{33}(\tau_2|\tau_1) ,\\ \nonumber
P_7\left(\tau_1,\tau_2;\lambda^{[\tau_1,\tau_2]}_7\right)&=&\frac{1}{4}[1-E_S(\tau_1)]E_L(\tau_1) 
[p_{33}(\tau_2|\tau_1)+p_{34}(\tau_2|\tau_1)] ,\\ \nonumber
P_8\left(\tau_1,\tau_2;\lambda^{[\tau_1,\tau_2]}_8\right)&=&
\frac{1}{4}E_S(\tau_1)E_L(\tau_1)p_{43}(\tau_2|\tau_1) ,\\ \nonumber
P_9\left(\tau_1,\tau_2;\lambda^{[\tau_1,\tau_2]}_9\right)&=&
\frac{1}{4}E_S(\tau_1)E_L(\tau_1)p_{22}(\tau_2|\tau_1) ,\\ \nonumber
P_{10}\left(\tau_1,\tau_2;\lambda^{[\tau_1,\tau_2]}_{10}\right)&=&
P_{14}\left(\tau_1,\tau_2;\lambda^{[\tau_1,\tau_2]}_{14}\right)=
\frac{1}{4}E_L(\tau_1)[1-E_S(\tau_2)] ,\\ \nonumber
P_{11}\left(\tau_1,\tau_2;\lambda^{[\tau_1,\tau_2]}_{11}\right)&=&
\frac{1}{4}E_S(\tau_1)[1-E_L(\tau_1)]
[p_{21}(\tau_2|\tau_1)+p_{22}(\tau_2|\tau_1)] ,\\ \nonumber
P_{12}\left(\tau_1,\tau_2;\lambda^{[\tau_1,\tau_2]}_{12}\right)&=&
\frac{1}{4}E_S(\tau_1)E_L(\tau_1)p_{12}(\tau_2|\tau_1) ,\\ \nonumber
P_{13}\left(\tau_1,\tau_2;\lambda^{[\tau_1,\tau_2]}_{13}\right)&=&
\frac{1}{4}E_S(\tau_1)E_L(\tau_1)p_{11}(\tau_2|\tau_1) ,\\ \nonumber
P_{15}\left(\tau_1,\tau_2;\lambda^{[\tau_1,\tau_2]}_{15}\right)&=&
\frac{1}{4}E_S(\tau_1)[1-E_L(\tau_1)]  
[p_{11}(\tau_2|\tau_1)+p_{12}(\tau_2|\tau_1)] ,\\ \nonumber
P_{16}\left(\tau_1,\tau_2;\lambda^{[\tau_1,\tau_2]}_{16}\right)&=&
\frac{1}{4}E_S(\tau_1)E_L(\tau_1)p_{21}(\tau_2|\tau_1) ,\\ \nonumber
P_{17}\left(\tau_1,\tau_2;\lambda^{[\tau_1,\tau_2]}_{17}\right)&=&
\frac{1}{2}[1-E_S(\tau_1)][1-E_L(\tau_2)] , \\ \nonumber
P_{18}\left(\tau_1,\tau_2;\lambda^{[\tau_1,\tau_2]}_{18}\right)&=&
\frac{1}{2}[1-E_L(\tau_1)][1-E_S(\tau_2)] . \nonumber
\end{eqnarray} 

The description of eqs.~(\ref{ourhv}) and (\ref{ourhv2}) and table
\ref{corrt} corresponds to the most general hidden--variable theory.
Actually, the local realistic probabilities of eqs.~(\ref{setprob0}), (\ref{setprob})
must be interpreted by means of equations like (\ref{non-det}),
namely they contain elements of randomness related both to apparata
hidden--variables and, in general, to other unknown mechanisms: 
$P_i(\tau_1,\tau_2;\lambda^{[\tau_1,\tau_2]}_i)=
P_i(\tau_1,\tau_2|\lambda^{[\tau_1,\tau_2]}_i)\rho(\lambda^{[\tau_1,\tau_2]}_i)$,
where $0\leq P_i(\tau_1,\tau_2|\lambda^{[\tau_1,\tau_2]}_i)\leq 1$.

When $\tau_1=\tau_2=0$, only the probabilities for the four states of table~\ref{corr0}:
\beq
P_1\left(0,0;\lambda^{[0,0]}_1\right)=P_5\left(0,0;\lambda^{[0,0]}_5\right)=
P_9\left(0,0;\lambda^{[0,0]}_9\right)=P_{13}\left(0,0;\lambda^{[0,0]}_{13}\right)=\frac{1}{4} ,
\eeq
are non--vanishing. Moreover, for $\tau_1=\tau_2\equiv \tau \neq0$, four probabilities of our
set are still zero:
\beq
P_4\left(\tau,\tau;\lambda^{[\tau,\tau]}_4\right)=P_8\left(\tau,\tau;\lambda^{[\tau,\tau]}_8\right)=
P_{12}\left(\tau,\tau;\lambda^{[\tau,\tau]}_{12}\right)=
P_{16}\left(\tau,\tau;\lambda^{[\tau,\tau]}_{16}\right)=0 ,
\eeq
because of the requirement of perfect anti--correlation on strangeness at equal times.

Consider now the contribution to $P_1(\tau,\tau;\lambda^{[\tau,\tau]}_1)$
coming from the transitions $K_1(0)\to K_1(\tau)$ on the left and
$K_4(0)\to K_4(\tau)$ on the right. It can be written in the following two
equivalent ways: 1) the probability that the left going kaon is
created in the state $K_1$ and is then subject to the
transition $K_1(0)\to K_1(\tau)$ is $p_{11}(\tau|0)/4$; in order to obtain the
required probability we have to multiply this quantity by the probability
$E_L(\tau)$ that the right going kaon at time $\tau$, that is correlated with
the left going $K_1$, is an undecayed $K_4$; 2) the probability that
the right going kaon is created in the state $K_4$ and is then
subject to the transition $K_4(0)\to K_4(\tau)$ is
$p_{44}(\tau|0)/4$; to obtain the required
probability we have to multiply this quantity by the probability
$E_S(\tau)$ that the left going kaon at time $\tau$ is an undecayed $K_1$.
Therefore, the following equality is valid:
\beq
p_{11}(\tau|0)E_L(\tau)=p_{44}(\tau|0)E_S(\tau) ,
\eeq
and from eq.~(\ref{matrix}) we obtain that it is verified only when $\delta(\tau)\equiv 0$.
This property can also be proved starting from the other probabilities of 
eqs.~(\ref{setprob0}), (\ref{setprob})
which are non--vanishing when $\tau_1=\tau_2$.

The independent observables relevant for the problem
[given, in quantum mechanics, by eqs.~(\ref{kkb})--(\ref{sl})] can be written, 
within local realism, as follows:
\begin{eqnarray}
\label{lr1}
P_{LR}[K^0(\tau_1),\bar{K}^0(\tau_2)]&\equiv & P_1\left(\tau_1,\tau_2;\lambda^{[\tau_1,\tau_2]}_1\right)+
P_9\left(\tau_1,\tau_2;\lambda^{[\tau_1,\tau_2]}_9\right) \\ \nonumber
&=&\frac{1}{4}E_S(\tau_1)E_L(\tau_1)[p_{22}(\tau_2|\tau_1)+p_{44}(\tau_2|\tau_1)] , \\ \nonumber
P_{LR}[\bar{K}^0(\tau_1),K^0(\tau_2)]&\equiv & P_5\left(\tau_1,\tau_2;\lambda^{[\tau_1,\tau_2]}_5\right)+
P_{13}\left(\tau_1,\tau_2;\lambda^{[\tau_1,\tau_2]}_{13}\right) \\ \nonumber
&=&\frac{1}{4}E_S(\tau_1)E_L(\tau_1)[p_{11}(\tau_2|\tau_1)+p_{33}(\tau_2|\tau_1)] , \\ \nonumber
P_{LR}[K^0(\tau_1),K^0(\tau_2)]&\equiv & P_4\left(\tau_1,\tau_2;\lambda^{[\tau_1,\tau_2]}_4\right)+
P_{12}\left(\tau_1,\tau_2;\lambda^{[\tau_1,\tau_2]}_{12}\right) \\ \nonumber
&=&\frac{1}{4}E_S(\tau_1)E_L(\tau_1)[p_{12}(\tau_2|\tau_1)+p_{34}(\tau_2|\tau_1)] , \\ \nonumber
P_{LR}[\bar{K}^0(\tau_1),\bar{K}^0(\tau_2)]&\equiv & 
P_8\left(\tau_1,\tau_2;\lambda^{[\tau_1,\tau_2]}_8\right)+
P_{16}\left(\tau_1,\tau_2;\lambda^{[\tau_1,\tau_2]}_{16}\right) \\ \nonumber
&=&\frac{1}{4}E_S(\tau_1)E_L(\tau_1)[p_{21}(\tau_2|\tau_1)+p_{43}(\tau_2|\tau_1)] , \\ \nonumber
P_{LR}[K_L(\tau_1),K_S(\tau_2)]&\equiv & P_9\left(\tau_1,\tau_2;\lambda^{[\tau_1,\tau_2]}_9\right)+
P_{12}\left(\tau_1,\tau_2;\lambda^{[\tau_1,\tau_2]}_{12}\right) \\ \nonumber
&&+P_{13}\left(\tau_1,\tau_2;\lambda^{[\tau_1,\tau_2]}_{13}\right)+
P_{16}\left(\tau_1,\tau_2;\lambda^{[\tau_1,\tau_2]}_{16}\right) \\ \nonumber
&=&\frac{1}{4}E_S(\tau_1)E_L(\tau_1)
[p_{11}(\tau_2|\tau_1)+p_{12}(\tau_2|\tau_1)+p_{21}(\tau_2|\tau_1)+p_{22}(\tau_2|\tau_1)] , \\ \nonumber
P_{LR}[K_S(\tau_1),K_L(\tau_2)]&\equiv & P_1\left(\tau_1,\tau_2;\lambda^{[\tau_1,\tau_2]}_1\right)+
P_4\left(\tau_1,\tau_2;\lambda^{[\tau_1,\tau_2]}_4\right) \\ \nonumber
&&+P_5\left(\tau_1,\tau_2;\lambda^{[\tau_1,\tau_2]}_5\right)+
P_8\left(\tau_1,\tau_2;\lambda^{[\tau_1,\tau_2]}_8\right) \\ \nonumber
&=&\frac{1}{4}E_S(\tau_1)E_L(\tau_1)
[p_{33}(\tau_2|\tau_1)+p_{34}(\tau_2|\tau_1)+p_{43}(\tau_2|\tau_1)+p_{44}(\tau_2|\tau_1)] .
\end{eqnarray}
The probabilities of eqs.~(\ref{skappa}) and (\ref{lkappa}) do not supply new 
information since they are not independent of the other ones just considered,
whereas eq.~(\ref{ss}), which ensures $CP$ conservation, was assumed,
in section \ref{localrealismkappa} and then in table~\ref{corrt}, 
when we introduced local realism for the two--kaon system.

In order to determine the observables of eq.~(\ref{lr1}), we now ask whether it
is possible to derive useful relations
among the $p_{ij}(\tau_2|\tau_1)$'s and the probabilities $p_{ij}(\tau|0)$
of matrix (\ref{matrix}). By introducing three--time probabilities:
\beq
\label{three}
p_{ijk}(\tau_2,\tau_1,0)=p_{ijk}(\tau_2,\tau_1|0)p_k(0)=
p_{ijk}(\tau_2|\tau_1,0)p_{jk}(\tau_1|0)p_k(0) ,
\eeq
and using the multiplication theorem, 
$p_{11}(\tau_2|\tau_1)$, $p_{12}(\tau_2|\tau_1)$, $p_{21}(\tau_2|\tau_1)$
and $p_{22}(\tau_2|\tau_1)$ can be written as follows:
\begin{eqnarray}
\label{p11d}
p_{11}(\tau_2,\tau_1)\equiv p_{11}(\tau_2|\tau_1)p_1(\tau_1)
&=&\frac{1}{4}\left[p_{111}(\tau_2,\tau_1|0)+p_{112}(\tau_2,\tau_1|0)\right] , \\
\label{p12d}
p_{12}(\tau_2,\tau_1)\equiv p_{12}(\tau_2|\tau_1)p_2(\tau_1)
&=&\frac{1}{4}\left[p_{121}(\tau_2,\tau_1|0)+p_{122}(\tau_2,\tau_1|0)\right] . \\
\label{p21d}
p_{21}(\tau_2,\tau_1)\equiv p_{21}(\tau_2|\tau_1)p_1(\tau_1)
&=&\frac{1}{4}\left[p_{211}(\tau_2,\tau_1|0)+p_{212}(\tau_2,\tau_1|0)\right] , \\
\label{p22d}
p_{22}(\tau_2,\tau_1)\equiv p_{22}(\tau_2|\tau_1)p_2(\tau_1)
&=&\frac{1}{4}\left[p_{221}(\tau_2,\tau_1|0)+p_{222}(\tau_2,\tau_1|0)\right] .
\end{eqnarray}
In the previous relations, 
$p_1(\tau_1)=[p_{11}(\tau_1)+p_{12}(\tau_1)]/4=E_S(\tau_1)/4$
($p_2(\tau_1)=[p_{21}(\tau_1)+p_{22}(\tau_1)]/4=E_S(\tau_1)/4$)
is the probability to observe a $K_1$ ($K_2$) along direction 1 at time $\tau_1$
(analogous relations are valid for $CP=-1$ states, and $p_3(\tau)=p_4(\tau)=E_L(\tau)/4$),
the $p_{jk}(\tau|0)$'s are given in eq.~(\ref{matrix}) with $\delta(\tau)\equiv 0$,
whereas $p_{ij}(\tau_2,\tau_1)$ denote standard
(namely non--conditional) two--times probabilities.
Moreover, $p_{ijk}(\tau_2,\tau_1,0)$ is the probability to have states $K_k$, $K_j$ and $K_i$
at times $0$, $\tau_1$ and $\tau_2$, respectively, $p_{ijk}(\tau_2,\tau_1|0)$
is the probability that at times $\tau_1$ and $\tau_2$ the states are 
$K_j$ and $K_i$, respectively, if the state at time $0$ was $K_k$, and,
finally, $p_{ijk}(\tau_2|\tau_1,0)$ is the probability of a $K_i$ at time 
$\tau_2$ conditional on the presence of a $K_k$ at time $0$ and a $K_j$ at $\tau_1$.
It is then clear that, in eqs.~(\ref{p11d})--(\ref{p22d}), the two--times probabilities 
$p_{ij}(\tau_2,\tau_1)$ are obtained by summing over the possible states 
appearing at time $\tau=0$.

Let us now consider probability $p_{11}(\tau_2|0)$. 
Introduce a time $\tau_1$ in the interval $[0,\tau_2]$: 
at instant $\tau_1$ the state can be
either a $K_1$ or a $K_2$, then the contributions to $p_{11}(\tau_2|0)$ come from
two transitions with different intermediate state. They are 
$K_1(0)\to K_1(\tau_1)\to K_1(\tau_2)$ and
$K_1(0)\to K_2(\tau_1)\to K_1(\tau_2)$, thus:
\beq
\label{p11}
p_{11}(\tau_2|0)=p_{111}(\tau_2,\tau_1|0)+p_{121}(\tau_2,\tau_1|0) ,
\eeq
for any $\tau_1\in [0,\tau_2]$.
Limiting again the discussion to probabilities relevant for the evolution of $CP=+1$ 
states, one obtains the remaining relations:
\begin{eqnarray}
\label{p12}
p_{12}(\tau_2|0)&=&p_{112}(\tau_2,\tau_1|0)+p_{122}(\tau_2,\tau_1|0) , \\
\label{p21}
p_{21}(\tau_2|0)&=&p_{211}(\tau_2,\tau_1|0)+p_{221}(\tau_2,\tau_1|0) , \\
\label{p22} 
p_{22}(\tau_2|0)&=&p_{212}(\tau_2,\tau_1|0)+p_{222}(\tau_2,\tau_1|0) . 
\end{eqnarray}

Now, the sum of two three--times probabilities corresponding to the same states at times
$0$ and $\tau_1$ but with different states at $\tau_2$ provides a known result; in fact:
\begin{eqnarray}
\label{cp11}
p_{111}(\tau_2|\tau_1,0)+p_{211}(\tau_2|\tau_1,0)&=&E_S(\tau_2-\tau_1) , \\
\label{cp12}
p_{112}(\tau_2|\tau_1,0)+p_{212}(\tau_2|\tau_1,0)&=&E_S(\tau_2-\tau_1) , \\
\label{cp21}
p_{121}(\tau_2|\tau_1,0)+p_{221}(\tau_2|\tau_1,0)&=&E_S(\tau_2-\tau_1) , \\
\label{cp22}
p_{122}(\tau_2|\tau_1,0)+p_{222}(\tau_2|\tau_1,0)&=&E_S(\tau_2-\tau_1) .
\end{eqnarray}
Each of these equalities accounts for the contributions to
transitions into final states $K_1$ and $K_2$ once the kaonic states
at times $0$ and $\tau_1$ are fixed: these probabilities equal
the probability $E_S(\tau_2-\tau_1)$ that a $CP=+1$ kaon does not decay during 
the time interval between $\tau_1$ and $\tau_2$.

By using the shorthand notation $p_{ijk}\equiv p_{ijk}(\tau_2,\tau_1|0)$, 
it follows from (\ref{matrix}) that the
above equations (\ref{p11})--(\ref{cp22}) can be written in the equivalent form:
\begin{eqnarray}
\label{primap}
p_{111}+p_{121}&=&p_{222}+p_{212}=E_S(\tau_2)Q_+(\tau_2)  \\
p_{112}+p_{122}&=&p_{221}+p_{211}=E_S(\tau_2)Q_-(\tau_2)  \\
p_{111}+p_{211}&=&p_{222}+p_{122}=E_S(\tau_2)Q_+(\tau_1)  \\
\label{quarta}
p_{112}+p_{212}&=&p_{221}+p_{121}=E_S(\tau_2)Q_-(\tau_1) .
\end{eqnarray}
These conditions on the 8 $CP=+1$ three--times probabilities 
supplies two system of equations:
\begin{equation}
\label{sistema1}
\left\{\begin{array}{l}
p_{111}+p_{121}=E_S(\tau_2)Q_+(\tau_2) \\
p_{221}+p_{211}=E_S(\tau_2)Q_-(\tau_2) \\
p_{111}+p_{211}=E_S(\tau_2)Q_+(\tau_1)
\end{array}\right.  ,
\end{equation}
\begin{equation}
\label{sistema2}
\left\{\begin{array}{l}
p_{222}+p_{212}=E_S(\tau_2)Q_+(\tau_2) \\
p_{112}+p_{122}=E_S(\tau_2)Q_-(\tau_2) \\
p_{222}+p_{122}=E_S(\tau_2)Q_+(\tau_1)
\end{array}\right.  ,
\end{equation}
each containing three independent conditions and four unknown probabilities.

From previous results one obtains:
\beq
\label{11-12}
p_{11}(\tau_2|\tau_1)+p_{12}(\tau_2|\tau_1)=\frac{1}{4}\left[
\frac{p_{111}+p_{112}}{p_1(\tau_1)}+\frac{p_{121}+p_{122}}{p_2(\tau_1)}\right]=
E_S(\tau_2-\tau_1)  .
\eeq
Analogously:
\beq
\label{altre}
p_{21}(\tau_2|\tau_1)+p_{22}(\tau_2|\tau_1)=
p_{11}(\tau_2|\tau_1)+p_{21}(\tau_2|\tau_1)
=p_{12}(\tau_2|\tau_1)+p_{22}(\tau_2|\tau_1)
=E_S(\tau_2-\tau_1) ,
\eeq
thus:
\begin{eqnarray}
p_{21}(\tau_2|\tau_1)&=&p_{12}(\tau_2|\tau_1) ,\\
\label{ultima}
p_{22}(\tau_2|\tau_1)&=&p_{11}(\tau_2|\tau_1) .
\end{eqnarray}
Exactly the same derivation can be repeated for the $CP=-1$ probabilities:
the relations valid in this case are obtained from (\ref{p11d})--(\ref{ultima})
simply by replacing $E_S$ with $E_L$ and
$1\to 3$, $2\to 4$ for the state indexes. Obviously,  
the normalization of the local realistic probabilities (\ref{setprob0}) and (\ref{setprob}),
$\sum_{i=1}^{18}P_i(\tau_1,\tau_2;\lambda^{[\tau_1,\tau_2]}_i)=1$,
is automatically ensured by the above results.

From eqs.~(\ref{lr1}) and previous analysis one thus obtains the following
expression for the observables within the local realistic approach:
\begin{eqnarray}
\label{lr00b}
P_{LR}[K^0(\tau_1),\bar{K}^0(\tau_2)]&=&P_{LR}[\bar{K}^0(\tau_1),K^0(\tau_2)]
=\displaystyle\frac{1}{8}[E_S(\tau_1)E_L(\tau_2)+ E_L(\tau_1)E_S(\tau_2)] \\
&&\times [1+A_{LR}(\tau_1,\tau_2)] , \nonumber \\
\label{lr00}
P_{LR}[K^0(\tau_1),K^0(\tau_2)]&=&P_{LR}[\bar{K}^0(\tau_1),\bar{K}^0(\tau_2)]
=\displaystyle\frac{1}{8}[E_S(\tau_1)E_L(\tau_2)+ E_L(\tau_1)E_S(\tau_2)] \\
&&\times [1-A_{LR}(\tau_1,\tau_2)] , \nonumber \\
\label{lrls}
P_{LR}[K_L(\tau_1),K_S(\tau_2)]&=&\displaystyle\frac{1}{4}E_L(\tau_1)E_S(\tau_2) ,  \\ 
\label{lrsl}
P_{LR}[K_S(\tau_1),K_L(\tau_2)]&=&\displaystyle\frac{1}{4}E_S(\tau_1)E_L(\tau_2) ,
\end{eqnarray}
written directly in terms of the asymmetry parameter [see definition (\ref{asimm})]:
\begin{eqnarray}
\label{lr-asimm}
A_{LR}(\tau_1,\tau_2)&=&2\frac{\left[p_{11}(\tau_2|\tau_1)+p_{33}(\tau_2|\tau_1)\right]}
{E_S(\tau_2-\tau_1)+E_L(\tau_2-\tau_1)}-1 \\
&=&2\frac{(p_{111}+p_{112})/E_S(\tau_1)+(p_{333}+p_{334})/E_L(\tau_1)}
{E_S(\tau_2-\tau_1)+E_L(\tau_2-\tau_1)}-1 , 
\nonumber
\end{eqnarray}
where, compatibly with constraints (\ref{sistema1}) and (\ref{sistema2}), 
the three--times probabilities can vary in the following intervals:
\begin{eqnarray}
\label{intervalli}
{\rm Max}\{0;Q_+(\tau_2)-Q_-(\tau_1)\} &\leq& 
\frac{p_{111}}{E_S(\tau_2)} ,\; \frac{p_{333}}{E_L(\tau_2)}\leq {\rm Min}\{Q_+(\tau_1);Q_+(\tau_2)\} , \\
{\rm Max}\{0;Q_-(\tau_1)-Q_+(\tau_2)\} &\leq& 
\frac{p_{112}}{E_S(\tau_2)} ,\; \frac{p_{334}}{E_L(\tau_2)}\leq {\rm Min}\{Q_-(\tau_1);Q_-(\tau_2)\} .
\nonumber
\end{eqnarray}

\section{Compatibility between local realism and quantum mechanics}
\label{comp}
From eqs.~(\ref{ourhv2}), (\ref{setprob0}), (\ref{setprob}), (\ref{11-12}), and (\ref{altre})
we obtain that local realism reproduces the single kaon quantum--mechanical 
expectation values:
\begin{eqnarray}
\label{comp-single}
P_{LR}[\bar{K}^0(\tau)]\equiv P_{QM}[\bar{K}^0(\tau)]&=&
\frac{1}{4}[E_S(\tau)+E_L(\tau)] , \\ 
P_{LR}[K^0(\tau)]\equiv P_{QM}[K^0(\tau)]&=&\frac{1}{4}[E_S(\tau)+E_L(\tau)] , \nonumber \\
P_{LR}[K_S(\tau)]\equiv P_{QM}[K_S(\tau)]&=&\frac{1}{2}E_S(\tau) ,  \nonumber \\
P_{LR}[K_L(\tau)]\equiv P_{QM}[K_L(\tau)]&=&\frac{1}{2}E_L(\tau) .  \nonumber
\end{eqnarray}
This result is of general validity \cite{Be64}: for all EPR--like particle pairs it is always possible
to take into account of the single particle observables by employing a local hidden--variable model.

Results (\ref{lrls}) and (\ref{lrsl}) reproduce the quantum--mechanical predictions
(\ref{ls}) and (\ref{sl}); it is easy to see that 
expectation values (\ref{skappa}) and (\ref{lkappa}) are obtained too.
The same conclusion would be true for the
joint observables (\ref{lr00b}) and (\ref{lr00})
involving $K_S$--$K_L$ mixing if the time--dependent local realistic 
asymmetry parameter had the same expression it has in quantum mechanics. Thus:
\beq
\label{lr-eq-qm}
{\rm Local\; Realism\; equivalent\; to\; Quantum\; Mechanics}\hspace{0.2cm}
\Longleftrightarrow
\hspace{0.2cm}A_{LR}(\tau_1,\tau_2)\equiv A_{QM}(\tau_1,\tau_2) .
\eeq
From eqs~(\ref{lr-asimm}) and (\ref{intervalli}) it follows that
the asymmetry corresponding to the most general local realistic theory 
satisfies the following inequality:
\beq
\label{interval}
2|Q_+(\tau_2)-Q_-(\tau_1)|-1\leq A_{LR}(\tau_1,\tau_2)\leq 1-2|Q_+(\tau_2)-Q_+(\tau_1)| .
\eeq

\begin{figure}[thb]
\begin{center}
\input{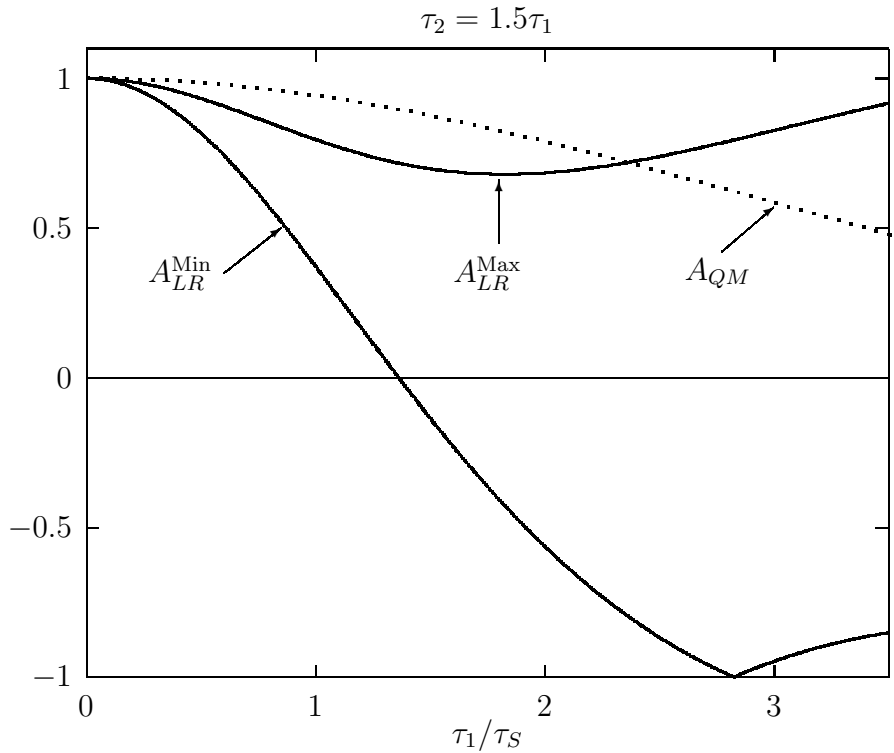}
\end{center}
\vskip 1.5mm
\caption{Local realistic and quantum--mechanical asymmetry parameters for $\tau_2=1.5\tau_1$
plotted vs $\tau_1/\tau_S$.}
\label{inc1}
\end{figure}

If one considers the special case in which $\tau_2=\tau_1\equiv \tau$, local realism
is compatible with quantum mechanics: in fact, 
$A_{LR}^{\rm Min}(\tau,\tau)\leq A_{QM}(\tau,\tau)=A_{LR}^{\rm Max}(\tau,\tau)\equiv 1$ for all times.
This is obvious, since within the class of local realistic theories supplying asymmetry
parameters in the interval (\ref{interval}), a model that reproduces the perfect anti--correlation
properties of the two--kaon state at equal times must exist.
When $\tau_1=0$, both descriptions supplies the same asymmetry: 
$A_{LR}(0,\tau)= A_{QM}(0,\tau)\equiv Q_+(\tau)-Q_-(\tau)$.
Another special case is when, for instance, $\tau_2=1.5\tau_1$: in this situation, 
the local realistic asymmetry does not satisfy
the compatibility requirement (\ref{lr-eq-qm}). This is depicted in figure~\ref{inc1}. 
The maximum values of the local realistic asymmetry stands below the 
quantum--mechanical ones for $0< \tau_1\lsim 2.3\tau_S$.
The largest incompatibility corresponds to $\tau_1\simeq 1.5\tau_S$, where
$[A_{QM}-A_{LR}^{\rm Max}]/A_{QM}\simeq 20$\%. 
In general, local realism and quantum mechanics are incompatible 
when $\tau_2=\alpha\tau_1$ with $\alpha> 1$. The degree of incompatibility
increases for increasing $\alpha$. For instance, when $\tau_2=2\tau_1\simeq 2.4\tau_S$, 
$A_{QM}$ is 27~\% larger than $A_{LR}^{\rm Max}$. The large differences among
quantum--mechanical and local realistic predictions justify our approach,
which neglected $CP$ violation.

However, it is important to stress the following restriction concerning the 
choice (which must be at free will) of the detection times $\tau_1$ and $\tau_2$.
In order to satisfy the locality condition, namely to make sure that the measurement 
event on the right is causally disconnected from that on the left,
these events must be space--like separated. For a two--kaon system
in which the kaons fly back--to--back in the laboratory frame system,
this requirement corresponds to choose detection times which satisfy the
inequality \cite{DD95}:
\beq
\label{loc-exp}
1\leq \frac{\tau_2}{\tau_1}< \frac{1+v}{1-v}=1.55 ,
\eeq
where $v\simeq 0.22$ is the kaon velocity (in units of $c$) in the laboratory frame system.
Nevertheless, concerning the locality assumption, a loophole that is impossible to avoid
could allow, in principle (it is completely unknown, however, in which way), an information to reach
both devices at the instants of measurement, whatever the choice of these times is.
In fact, events in the overlap region of the two backward light--cones corresponding
to the measurements at $\tau_1$ and $\tau_2$ might be responsible for the choice of the
times $\tau_1$ and $\tau_2$ as well as for the experimental outcomes. If this were the actual case,
even for causally disconnected measurement events one could not infer that the non--occurrence
of action--at--a--distance implies locality. Thus, a non--local behaviour of microscopic 
phenomena could be still compatible with relativistic causality. 

An experiment that measured the asymmetry parameter
was performed by the CPLEAR collaboration at CERN \cite{Ap98}.
The $K^0\bar{K}^0$ pairs were produced by proton--antiproton
annihilation at rest, while the kaon strangeness was detected through 
kaon strong interactions with bound nucleons of absorber materials.
The data, corrected for a comparison
with pure quantum--mechanical predictions [eq.~(\ref{qmasymm})], are reported in table \ref{cplear}.
\begin{table}[t]
\begin{center}
\caption{Asymmetry parameter measured by CPLEAR collaboration \protect\cite{Ap98}.}
\label{cplear}
\begin{tabular}{c|c c c}
\mc {1}{c|}{Time difference: $\tau_2-\tau_1$} &
\mc {1}{c}{Experiment} &
\mc {1}{c}{Quantum Mechanics} & 
\mc {1}{c}{Local Realism} \\ \hline
$0$           &$0.88\pm 0.17$ &$1$    & $0.86\div 1$     \\
$1.37\tau_S$  &$0.56\pm 0.12$ &$0.64$ & $0.34\div 0.48$  \\
\end{tabular}   
\end{center} 
\end{table}
The temporal uncertainty of data is not considered here.
Asymmetry values compatible with local realism depend on the detection
times $\tau_1$ and $\tau_2$ separately: the CPLEAR set--up corresponds to 
the following corrected times: $\tau_1=\tau_2=0.55\tau_S$ when $\tau_2-\tau_1=0$
and $\tau_1=0.55\tau_S$, $\tau_2=1.92\tau_S$ when $\tau_2-\tau_1=1.37\tau_S$.
We notice that also in the second case the 
two observation events were space--like separated:
in fact, $\tau_2/\tau_1=3.5$ (also when uncorrected times
are considered), and, since the kaon velocity in the center of mass system
for $p\bar{p}\to K^0\bar{K}^0$
is $v\simeq 0.85$, condition (\ref{loc-exp}) gives $1\leq \tau_2/\tau_1< 12.2$.
It is evident from table \ref{cplear} that the data are in agreement, within one
standard deviation, with both quantum mechanics and local realism.
For a decisive test of local realistic theories more precise data are needed. 

In agreement with Bell's theorem, in this section we have seen that
local realism contradicts some statistical
predictions of quantum mechanics concerning the evolution of the 
two--neutral--kaon system. Local realism has already been tested against quantum mechanics
(by employing Bell--type inequalities)
in optics and atomic physics: neglecting existing loopholes,
apart form some irrelevant exception,
all the experimental results 
revealed incompatible with the local realistic 
viewpoint and were in good agreement with quantum mechanics. 
For the two--kaon correlated system one avoids the detection loophole and,
in particular situations, the differences among the predictions
of local realism and quantum mechanics are so evident that a future measurement 
at the Frascati $\Phi$--factory
(say for $\tau_2=1.5\tau_1$, with $\tau_1$ around $1.5\tau_S$) should
be able to confirm one of the two pictures.

\section{On the possibility to test local realism with Bell's inequalities
for the two--neutral--kaon system}
\label{imposs}
When $CP$ non--conservation is taken into account, a Bell's inequality violated
by quantum mechanics has been derived in the special case of a gedanken experiment \cite{Uc97}.
Unfortunately, the magnitude of violation of this
inequality is very small, of the order of the $K^0$--$\bar{K}^0$
$CP$ violating parameter, $\epsilon$,
thus representing a problem from the experimental point of view.
A similar inequality, which is violated by a non--vanishing value of the direct $CP$
and $CPT$ violating parameter, $\epsilon^{\prime}$, is discussed in ref.~\cite{BF98}.
Moreover, experimental set--up exploiting $K_S$--$K_L$ regeneration processes have
also been proposed in order to formulate
Bell's inequalities that show incompatibilities with some statistical
predictions of quantum theory \cite{Eb93,DD95,Br99}. Unfortunately, 
in order to avoid a tiny violation of the inequalities that one obtains for thin
regenerators, this kind of Bell--type test requires large amount of regenerator materials.
Moreover, the test proposed in ref.~\cite{Eb93} can be performed only
at asymmetric $\Phi$--factories.

In ref.~\cite{Gh91} the authors concluded that, under the hypothesis of $CP$ conservation,
because of the specific properties of the kaon, it is impossible to test local realism 
by using Bell's inequalities, since whatever inequality one considers,
a violation by quantum--mechanical expectation values cannot be found.
In this section we consider again this question in order 
to prove how such a test is actually feasible with Wigner's inequalities \cite{Wi70}.
Moreover, in agreement with the discussion of ref.~\cite{Gi00}, we shall also show that
a Bell--test is possible when properly normalized observables and Clauser--Horne--Shimony--Holt's 
(CHSH's) inequalities \cite{CHSH69,CH74} are employed.

The two--kaon system presents some analogies but also a significant difference
compared to the case of the singlet state of two spin--$1/2$ particles
(\ref{spin0}). 
The (free) choice of the times $\tau_1$ and $\tau_2$ at which
a strangeness measurements on the kaon pair (\ref{qm2k}) is performed
is analogous to the (free) choice of the orientation along which the spin
is observed in the case of the singlet state (\ref{spin0}). 
If we consider the ideal limit in which the weak interaction eigenstates $K_S$ and $K_L$
are stable ($\Gamma_L=\Gamma_S=0$), quantum--mechanical kaon
probabilities (\ref{kkb}) and (\ref{kk})
have exactly the same expressions (proportional to $1\pm {\rm cos}\, \theta_{12}$)
of the spin--singlet case, provided one replaces the 
angle between the two spin analyzers with $\theta_{12}\equiv \Delta m(\tau_2-\tau_1)$. Then, 
a strangeness measurement on the two--kaon system is perfectly equivalent to a 
spin measurement on the singlet state (\ref{spin0}). 
It is then obvious that if the above hypothesis were
realized in Nature, one could find violations of Bell's inequalities of the same magnitude 
of the ones that characterize the spin system.

However, this hypothesis is far from being realistic, and the kaon joint probabilities
decreases with time because of the $K_S$ and $K_L$ weak decays. This leads to an important
difference with respect to the spin case. Because of the particular
values of the kaon lifetimes ($\Gamma_S$ and $\Gamma_L$) 
and of the quantity $\Delta m\equiv m_L-m_S$, which controls the 
quantum--mechanical interference term
(i.e., the $K_S$--$K_L$ mixing), ref.~\cite{Gh91} concluded  
that no choice of the detection times is able to show a violation, by quantum mechanics,
of Bell's inequalities. The authors of ref.~\cite{Gh91} reached this conclusion on the 
basis of CHSH's inequalities. 

Actually, the interplay between kaon exponential damping and strangeness oscillations
only makes it more difficult (but not impossible) a Bell--type test.
The reason of this behaviour lies in the very short $K_S$ lifetime 
($\tau_S$) compared with the typical time ($2\pi/\Delta m\simeq 13 \tau_S$) of the strangeness
oscillations. The situation would be different (namely the discrimination between 
quantum mechanics and local realistic theories would be easier)
if one treated the $B^0$--$\bar{B}^0$ system (this is due to the fact that
the states analogous to $K_S$ and $K_L$ for the $B^0$--meson have the same
lifetime). 

We start discussing Wigner's inequalities. We must recall that 
these inequalities
are derivable for deterministic theories only, therefore
they are less general than CHSH's.
Let us consider the following Wigner's inequality involving $K_S$--$K_L$
mixing:
\beq
\label{wigin}
P_{LR}[\bar{K}^0(\tau_1),\bar{K}^0(\tau_2)]\leq P_{LR}[\bar{K}^0(\tau_1),\bar{K}^0(\tau_3)]+
P_{LR}[\bar{K}^0(\tau_3),\bar{K}^0(\tau_2)] ,
\eeq
where $\tau_1\leq \tau_3\leq \tau_2$. It has been written for $\bar{K}^0\bar{K}^0$ 
joint detection since $\bar{K}^0$ states
are easier to detect than $K^0$ states. Obviously, the same conclusions
that we shall obtain in the following
are valid for the inequality corresponding to $K^0K^0$ detection. Inequalities
that contain $K^0\bar{K}^0$ joint probabilities turn out to be useless
for Bell-type tests. 

\begin{figure}[thb]
\begin{center}
\input{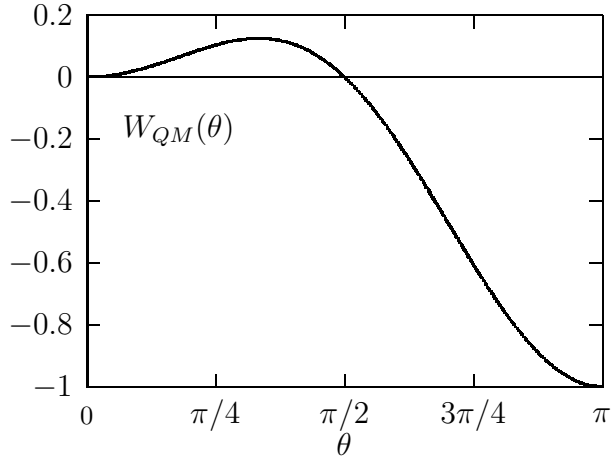}
\end{center}
\vskip 1.5mm
\caption{Violation of Wigner's inequality (\ref{wigin1}) for the spin--singlet state
(\ref{spin0}). The function $W_{QM}(\theta)\equiv P_{QM}(s_a=-,s_b=-)-P_{QM}(s_a=-,s_c=-)-
P_{QM}(s_c=-,s_b=-)$ is plotted versus $\theta$
($\theta\equiv \theta_{ab}/2=\theta_{ac}=\theta_{cb}$). The inequality
is violated by quantum mechanics when $\theta$ is in the interval $[0,\pi/2]$.}
\label{wig}
\end{figure}

In the limit $\Gamma_S=\Gamma_L=0$, eq.~(\ref{wigin})
reduces to the analogous inequality for the spin--singlet case:
\beq
\label{wigin1}
P_{LR}(s_a=-,s_b=-)\leq P_{LR}(s_a=-,s_c=-)+P_{LR}(s_c=-,s_b=-) .
\eeq 
Since:
\beq
P_{QM}(s_{\alpha}=-,s_{\beta}=-)=\frac{1}{4}(1-{\rm cos}\, \theta_{\alpha \beta}) ,
\eeq
$\theta_{\alpha \beta}$ being the angle between the spin measurement
directions characterized by the unitary vectors $\vec \alpha$ and $\vec \beta$,
inequality (\ref{wigin1}) is violated by quantum mechanics
when one chooses $\theta_{ab}=2\theta_{ac}=2\theta_{cb}\equiv 2\theta$ with
$\theta$ in the interval $[0,\pi/2]$ (see figure \ref{wig}). The greatest violation
of eq.~(\ref{wigin1}) ($0.375>0.250$) is for $\theta=\pi/3$ and is significant, since
it corresponds to $P_{QM}(s_a=-,s_b=-)=0.375$ and
$P_{QM}(s_a=-,s_c=-)=P_{QM}(s_c=-,s_b=-)=0.125$.

\begin{figure}[thb]
\begin{center}
\input{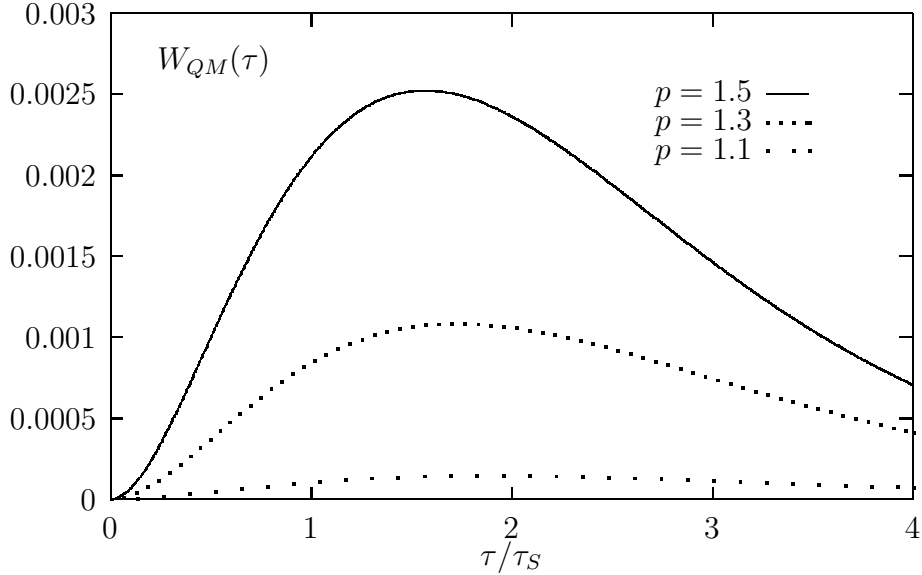}
\end{center}
\vskip 1.5mm
\caption{Violation of Wigner's inequality (\ref{wigin}) 
for $\tau_1=\tau_2/p=2\tau_3/(p+1)\equiv \tau$. The function $W_{QM}(\tau)\equiv
P_{QM}[\bar{K}^0(\tau_1),\bar{K}^0(\tau_2)]- P_{QM}[\bar{K}^0(\tau_1),\bar{K}^0(\tau_3)]-
P_{QM}[\bar{K}^0(\tau_3),\bar{K}^0(\tau_2)]$ is plotted versus $\tau/\tau_S$.
From the top to the bottom the curves correspond to 
$p=1.5$, $1.3$, and $1.1$, respectively.}
\label{wig2}
\end{figure}

Coming back to the two--kaon system in the real case with $\Gamma_S/\Gamma_L\simeq 579$,
we must require the three detection times of inequality (\ref{wigin}) to 
satisfy restriction (\ref{loc-exp}), dictated by the necessity
to avoid any causal connection between the measurements
that could be present if the two observation events would not be space--like separated.
By introducing the relation:
\beq
\tau_2-\tau_1=2(\tau_3-\tau_1)=2(\tau_2-\tau_3) \equiv (p-1)\tau 
\eeq
among the observation times and choosing $\tau_1=\tau$, one obtains
$\tau_2=p\tau$ and $\tau_3=(p+1)\tau/2$, and the locality 
requirement (\ref{loc-exp}) is fulfilled when $1\leq p< 1.55$. Apart form 
the case with $p=1$, for all values of $p$ in this range, 
inequality (\ref{wigin}) is incompatible with quantum mechanics at small $\tau$:
in figure \ref{wig2} this is shown for $p=1.5$, $1.3$, and $1.1$. 
In the case with $p=1.5$, the largest violation 
of eq.~(\ref{wigin}) ($0.0051>0.0026$) corresponds to 
$\tau \simeq 1.6\tau_S$, $P_{QM}[\bar{K}^0(\tau_1),\bar{K}^0(\tau_2)]=0.0051$,
$P_{QM}[\bar{K}^0(\tau_1),\bar{K}^0(\tau_3)]=0.0016$ and
$P_{QM}[\bar{K}^0(\tau_3),\bar{K}^0(\tau_2)]=0.0010$. 
An experimental test of a tiny violation like this one
is difficult to perform since, relatively to the typical experimental accuracy, 
the function $W_{QM}(\tau)$ of figure~\ref{wig2} is too close to 0 in the 
region of maximum violation. 
In particular, a precision like that of the CPLEAR
experiment (see table \ref{cplear}) is insufficient to achieve this purpose, since
$W(\tau)$ would be measured with an error larger than the maximum violation shown in
figure \ref{wig2}.

When one considers joint probabilities normalized to undecayed kaon pairs:
\begin{eqnarray}
\label{replace-prob}
P[\bar{K}^0(\tau),\bar{K}^0(\tau')]\to 
P^{\rm ren}[\bar{K}^0(\tau),\bar{K}^0(\tau')]&\equiv&
\frac{P[\bar{K}^0(\tau),\bar{K}^0(\tau')]}{P[-(\tau),-(\tau')]} \\
&=&\frac{1}{4}[1-A(\tau,\tau')] , \nonumber
\end{eqnarray}
since these quantities are less damped than the original ones,
a Bell--type test can be performed also with CHSH's inequalities.
In the previous equation, the probability that at times $\tau$ (on the left) 
and $\tau'$ (on the right) both kaons are undecayed is:
\begin{eqnarray}
P[-(\tau),-(\tau')]&=&P[\bar{K}^0(\tau),\bar{K}^0(\tau')]+
P[\bar{K}^0(\tau),K^0(\tau')]+P[K^0(\tau),\bar{K}^0(\tau')] \\
&&+P[K^0(\tau),K^0(\tau')]=
\frac{1}{2}[E_S(\tau)E_L(\tau')+E_L(\tau)E_S(\tau')] , \nonumber
\end{eqnarray}
the last equality being valid both in the local realistic description
[eqs.~(\ref{lr00b}), (\ref{lr00})] and in
quantum mechanics [eqs.~(\ref{kkb}), eqs.~(\ref{kk})], since it is
independent of the $K^0$--$\bar{K}^0$ oscillations.
The same derivation that supplies the CHSH's inequality in the unrenormalized
case can be applied to the renormalized
observables of eq.~(\ref{replace-prob}). By introducing four detection times
($\tau_1$ and $\tau_2$ for the left going meson, 
$\tau_3$ and $\tau_4$ for the right going meson),
the CHSH's inequality for strangeness $-1$ detection is then:
\beq
\label{chsh}
-1\leq S_{LR}(\tau_1,\tau_2,\tau_3,\tau_4)\leq 0 ,
\eeq
with:
\begin{eqnarray}
\label{chsh1}
S_{LR}(\tau_1,\tau_2,\tau_3,\tau_4)&\equiv&
P_{LR}^{\rm ren}[\bar{K}^0(\tau_1),\bar{K}^0(\tau_3)]-
P_{LR}^{\rm ren}[\bar{K}^0(\tau_1),\bar{K}^0(\tau_4)] 
+P_{LR}^{\rm ren}[\bar{K}^0(\tau_2),\bar{K}^0(\tau_3)] \\
&&+P_{LR}^{\rm ren}[\bar{K}^0(\tau_2),\bar{K}^0(\tau_4)]-
P_{LR}^{\rm ren}[\bar{K}^0(\tau_2)]-P_{LR}^{\rm ren}[\bar{K}^0(\tau_3)] , \nonumber
\end{eqnarray}
where the single meson observables are given by [see eq.~(\ref{comp-single})]:
\beq
P_{LR}^{\rm ren}[\bar{K}^0(\tau)]\equiv \frac{P_{LR}[\bar{K}^0(\tau)]}
{P_{LR}[-(\tau)]}=\frac{1}{2} .
\eeq
Consider the special case in which the four times are related by:
\beq
\label{timerel}
\tau_3-\tau_1=\tau_2-\tau_3=\tau_4-\tau_2=\frac{1}{3}(\tau_4-\tau_1)\equiv \tau .
\eeq
Thus, in quantum mechanics quantity (\ref{chsh1}) reduces 
to [see eq.~(\ref{replace-prob})]:
\beq
\label{chsh2}
S_{QM}(\tau)=\frac{1}{4}\left[2-3A_{QM}(\tau)+A_{QM}(3\tau)\right]-1 . 
\eeq

\begin{figure}[t]
\begin{center}
\input{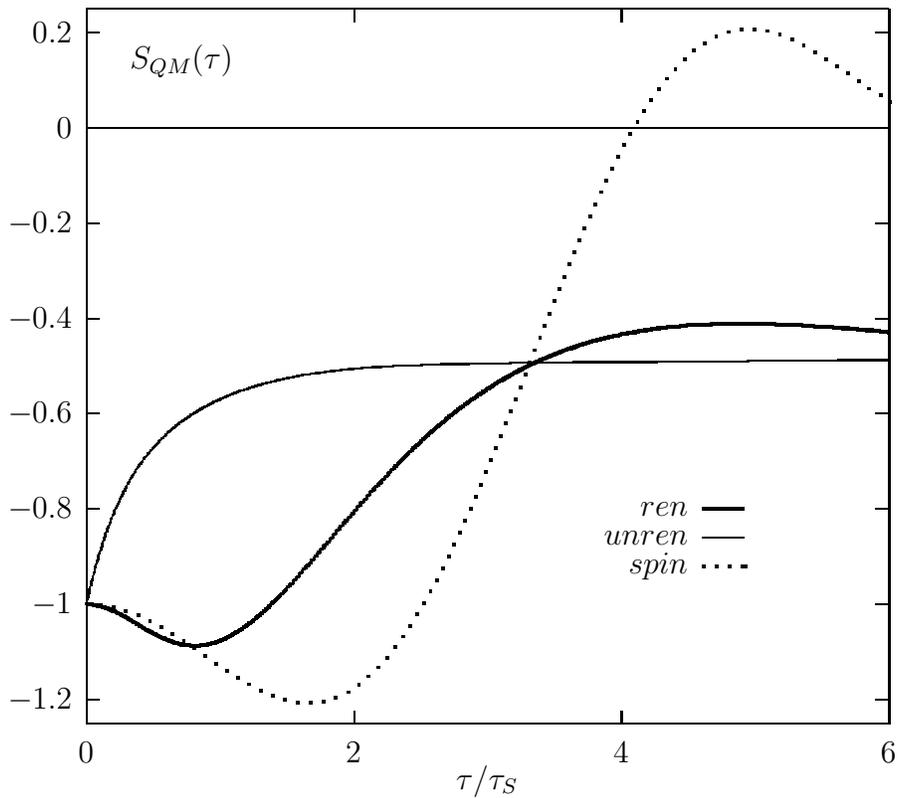}
\end{center}
\vskip 1.5mm
\caption{Violation of CHSH's inequality (\ref{chsh})
for $\tau_1/p=\tau_2/(p+2)=\tau_3/(p+1)=\tau_4/(p+3)\equiv \tau$.
The function $S_{QM}$ of eq.~(\ref{chsh2}) is plotted versus $\tau$.
The curve {\it unren} corresponds to the case employing unrenormalized
probabilities and $p=1$: the inequality is not violated by quantum--mechanical
observables. Values of $p$ compatible with the locality assumption
($p>5.45$) provide unrenormalized functions $S_{QM}$ more damped. Both curves valid
in the limit $\Gamma_S=\Gamma_L=0$ ({\it spin}) and in the real case of
unstable kaons with probabilities normalized to undecayed kaons ({\it ren})
violate CHSH's inequality and are independent of $p$.}
\label{chshfig}
\end{figure}

If we choose $\tau_1\equiv \tau$, the other times become:
$\tau_2=3\tau$, $\tau_3=2\tau$ and $\tau_4=4\tau$, and,
in the limit of stable kaons, both side of inequality
(\ref{chsh}) are violated by quantum mechanics in periodical intervals of $\tau$
(see curve marked {\it spin} in figure \ref{chshfig}): 
the largest violations are: $-1.21<-1$, $0.21>0$.

As far as the real case for kaons is considered, quantum mechanics does not
violate inequality (\ref{chsh}) when unrenormalized expectation values
are used (see curve {\it unren} in figure \ref{chshfig}). The conclusion is
different once one employs probabilities normalized 
to undecayed kaon pairs: as it is shown
in figure \ref{chshfig} (curve {\it ren}), for $0< \tau\lsim 1.4\tau_S$
quantum--mechanical expectation values are incompatible with the left hand side of 
inequality (\ref{chsh}). The largest violation of the inequality 
($-1.087<-1$) corresponds to $\tau\simeq 0.81\tau_S$ and 
$P_{QM}^{\rm ren}[\bar{K}^0(\tau_1),\bar{K}^0(\tau_3)]\simeq 0.036$,
$P_{QM}^{\rm ren}[\bar{K}^0(\tau_1),\bar{K}^0(\tau_4)]\simeq 0.195$.

With the previous choice of the four detection times the locality condition 
(\ref{loc-exp}) is not satisfied, since: $\tau_4/\tau_1=4>1.55$. In order to
fulfil this requirement when relation (\ref{timerel}) is used, one can introduce
times $\tau_1=p\tau$, $\tau_2=(p+2)\tau$, $\tau_3=(p+1)\tau$ and $\tau_4=(p+3)\tau$
($p\geq 0$) and require $\tau_4/\tau_1=(p+3)/p< 1.55$, thus $p> 5.45$. 
However, since the renormalized quantum--mechanical probabilities
only depend on the difference between the observation times
[see eqs.~(\ref{replace-prob}), (\ref{qmasymm})],
the result {\it ren} of figure \ref{chshfig}
is independent of $p$, and the locality condition is satisfied.
Thus, experimentally one could choose to use, for instance,
$p=6$, namely $\tau_1=6\tau$, $\tau_2=8\tau$, $\tau_1=7\tau$, 
$\tau_4=9\tau$, and the largest violation of the inequality would be again
for $\tau\simeq 0.81\tau_S$. However, as $p$ increases, even if the renormalized
probabilities are unchanged, the strangeness
detection becomes more and more difficult, because of the kaon decays, thus 
small $p$ are preferable.
Also the curve corresponding to the limit $\Gamma_S=\Gamma_L=0$ is the same for any $p$.
Of course, the curve corresponding to the inequality
that makes use of unrenormalized probabilities depends on $p$, but this case 
is not interesting since it cannot be used
for a discriminating test whatever the choice of $p$ is. 

Assuming the same relative error in the measurement of all joint 
probabilities appearing in
eq.~(\ref{chsh1}) and disregarding the uncertainties on the single kaon detection,
the need for an error on $S_{\rm Exp}$ {\it much smaller} than the maximum violation
(0.087) of figure \ref{chshfig} is satisfied if the joint probabilities
of (\ref{chsh1}) can be determined with an accuracy 
$\delta P^{\rm ren}_{\rm Exp}/P^{\rm ren}_{\rm Exp}<< 40$\%. The CPLEAR data did not 
fulfil this condition. The experimental accuracy required to perform a conclusive test
of local realism with CHSH's inequality (\ref{chsh}) is of the same order of magnitude
of that needed in the use of Wigner's inequality (\ref{wigin}). 
To give an idea of the comparison between the potentialities of
a test with Bell's inequalities and a test through 
the measurement of the asymmetry parameter, let us consider the following 
hypothetical case in the situation of figure \ref{inc1}
with $\tau_2=1.5\tau_1\simeq 2.3\tau_S$.
By assuming a relative precision on the measurement of $\bar{K}^0\bar{K}^0$ pairs five
times better than the one for $K^0\bar{K}^0$ detection, the requirement 
$\delta P^{\rm ren}_{\rm Exp}/P^{\rm ren}_{\rm Exp}<< 40$\% 
for the joint observables of (\ref{chsh1}) corresponds to
an accuracy on the asymmetry, $\delta A_{\rm Exp}/A_{\rm Exp}<< 20$\%, 
that would allow a clear test of local realism.

\section{Conclusions}
\label{concl}
In agreement with Bell's theorem, in this paper we have shown that
quantum mechanics for the two--neutral--kaon system cannot be completed by a
theory which is both local and realistic: the separability assumed 
in Bell's local realistic theories for the joint probabilities contradicts the
non--separability of quantum entangled states. Although both the locality condition
and the realistic viewpoint seem reasonable, they are not prescribed by any first
principle. 
Any local realistic approach is only able to reproduce the non--paradoxical
predictions of quantum mechanics like the
perfect anti--correlations in strangeness and $CP$ and the single meson observables.
On this point it is important to recall that
the authors of ref.~\cite{Gr90} showed how for entangled systems of three or more particles
the incompatibility between local realism and quantum mechanics is even deeper:
in fact, for these systems, a contradiction already arises at the level of 
perfectly correlated quantum states,
the premises of local realism being in conflict with the non--statistical 
predictions of quantum mechanics. For EPR's pairs,
maintaining the realistic viewpoint, in order to reproduce the
prediction of quantum mechanics (which, up to now, have been strongly
supported by experimental evidence), one is forced to consider
as a real fact of Nature a non--local behaviour of microscopic phenomena.

In the present paper,
the incompatibility proof among quantum mechanics and local realistic models
has been carried out by employing two different
approaches. We started discussing the variability of the
expectation values deduced from the general premises
concerning locality and realism. The realistic states have been interpreted
within the widest class of hidden--variable models. 
As far as the process $e^+e^-\to \phi \to K^0\bar{K}^0$ is considered,
under particular conditions
for the experimental parameters (the detection times), the discrepancies 
among quantum mechanics and local realistic models for the 
time--dependent asymmetry are not less than
20\%. The data collected by the CPLEAR collaboration 
(which used the reaction $p \bar{p}\to K^0\bar{K}^0$ ) do not allow
for conclusive answers concerning a refutation of
local realism: these data are compatible
not only with quantum--mechanical asymmetries, but
with the range of variability of local realistic predictions.
Therefore, a decisive test of local realism needs for more precise data.

The other approach we followed in this paper makes use of Bell--like inequalities
involving $K_S$--$K_L$ mixing.
Contrary to what is generally believed in the literature, we have shown that a Bell--type
test is feasible at a $\Phi$--factory, both with Wigner's and
(once  probabilities normalized to undecayed kaons are used)
CHSH's inequalities. 
For an experiment at a $\Phi$--factory, the degree of
inconsistency between quantum mechanics and local realism shown by a Bell test
is of the same order of magnitude of that obtained in the first part of the paper
through the comparison of the asymmetry parameters. 

Concluding, by employing an experimental accuracy for joint kaon detection
considerably higher than that corresponding to the CPLEAR measurement,
a decisive test of local realism vs quantum mechanics
both with and without the use of Bell's inequalities will be feasible in the
future at the Frascati $\Phi$--factory.

\acknowledgments
We are grateful to Albert Bramon for many valuable discussions.
One of us (G.G.) acknowledge financial support by the EEC through 
TMR Contract CEE--0169.

\vfill\eject
\end{document}